\documentclass[journal]{IEEEtran}
\usepackage{amsmath}
\usepackage[T1]{fontenc}
\usepackage[latin9]{luainputenc}
\usepackage{amsbsy}
\usepackage{amstext}
\usepackage{amssymb}
\usepackage{graphicx}
\usepackage{subfigure}
\usepackage{esint}
\usepackage{amsthm}
\usepackage{float}
\usepackage{multirow}
\usepackage{stfloats}
\usepackage[noend]{algpseudocode}
\usepackage{algorithmicx,algorithm}
\usepackage{booktabs}

\usepackage[unicode=true,
 bookmarks=true,bookmarksnumbered=true,bookmarksopen=true,bookmarksopenlevel=1,
 breaklinks=false,pdfborder={0 0 0},pdfborderstyle={},backref=false,colorlinks=false]
 {hyperref}
 
\begin{document}
\title{A Vehicles Control Model to Alleviate Traffic Instability}
\author{Jiancheng Fang,
		Yu Xiang,~\IEEEmembership{Member,~IEEE,}
        Yu Huang,
        Yilong Cui,
        Wenyong Wang,~\IEEEmembership{Member,~IEEE,}
\thanks{Yu Xiang University of Electronic Science and Technology of China, Chengdu 611731, China, e-mail:jcxiang@uestc.edu.cn}
\thanks{Jiancheng Fang is with School of Computer Science and Engineering, University of Electronic Science and Technology of China, Chengdu 611731, China}
\thanks{Manuscript received April XX, XXXX; revised August XX, XXXX.}}

\markboth{Journal of \LaTeX\ Class Files,~Vol.~14, No.~8, August~2015}%
{Shell \MakeLowercase{\textit{et al.}}: Bare Demo of IEEEtran.cls for IEEE Journals}

\maketitle

\begin{abstract}
While bringing convenience to people, the growing number of vehicles on road already cause inevitable traffic congestion. Some traffic congestion happen with observable reasons, but others occur without apparent reasons or bottlenecks, which referred to as phantom jams, are caused by traditional vehicle following model. In order to alleviate the traffic instability caused by phantom jam, several models have been proposed with the development of  intelligent transportation system (ITS). these have been proved to be able to suppress traffic instability in the ideal situation. But in road scenarios, uncertainties of vehicle state measurements and time delay caused by on-board sensors, inter-vehicle communications and control system of vehicles will affect the performance of the existing models severely, and cannot be ignored. In this paper, a novel predictable bilateral control model-PBCM, which consists of best estimation and state prediction is proposed to determine accurate acceleration values of the host vehicle in traffic flow to alleviate traffic instability. Theoretical analysis and simulation results show that our model could reduce the influence of the measurement errors and the delay caused by communication and control system effectively, control the state of the vehicles in traffic flow accurately, thus achieve the goal of restrain the instability of traffic flow.
\end{abstract}


\IEEEpeerreviewmaketitle

\section{Introduction}
\IEEEPARstart{T}{raffic} congestion refers to the phenomenon that traffic with slower vehicle velocity and longer vehicle queues. According to the researchers, traffic congestion has caused a number of negative impacts to human and society. For example, the economic loss it caused to each U.S. driver was \$1445 in 2017 and RMB1200 in Beijing, China in 2019, due to the time wasted on roads\cite{ref1,ref2}. Also, while congestion, wasted fuel increasing greenhouse gas such as vehiclebon dioxide emissions owing to increased idling, acceleration and braking \cite{ref3}. Furthermore, $80\%$ amount of drivers have reported felt anger, aggression or stress while driving with traffic congestion, thus reduced their health \cite{ref1}. Some traffic congestion happens with obvious reasons, such as bad weather, vehicle accident, road construction etc., but some occurs without any apparent reasons or bottlenecks. This kind of traffic congestion is also referred to as phantom jam that often appears in urban areas or on freeways and is first demonstrated through aerial photographs in \cite{ref4}. Sugiyama et al. showed that while road's capacity was reached with increasing demand, the traffic flow becomes unstable that even a small perturbation such as abrupt steering maneuver would cause phantom jam \cite{ref5}.

A number of models and methods were already proposed to analyze the traffic congestion in recent years. Fluid-dynamical models, also called macroscopic models were applied to traffic flow. B. Seibold et.al used classical Payne-Whitham model and the in homogeneous Aw-Rascle-Zhang model to demonstrate the phase transitions between stable and jamiton-dominated of traffic flow \cite{ref7}. N Malvin et. al used grid-based model instead of Lagrangian particle method to solve Payne-Whitham equation \cite{ref6}. Percolation theory already has been applied to study the critical percolation properties within traffic congestion in a city to find the possible equilibrium point between traffic supply and demand \cite{ref8}. Economic theories also introduced to analyze traffic congestion. Because roads in most places are free at the point of usage, there is little financial incentive for drivers not to over-use them, up to the point where traffic collapses into a jam, when demand becomes limited by opportunity cost \cite{ref9,ref10}. Micro-economic modelling approach with prospect theory and regret theory was explored in \cite{ref11} to describe observed microscopic and macroscopic traffic conditions and evaluate the behavioral heterogeneities' impact on traffic mobility and safety. Most of these models are not focused on phantom jam, and more importantly, those global mathematical models with relatively high computational complexity, thus make them hard to use under road circumstances.

Recently, ITS proved to be an effective way to mitigate the problem of traffic congestion \cite{ref12}. With ITS, drivers/autonomous vehicles can obtain real time traffic information along the road and other vehicles' current state (e.g. position, velocity, acceleration, etc.), thus facilitate their decision making on travel planning and navigation. These changes in control decisions shift the pattern of overall traffic demand and the properties of the traffic flow, thereby mitigating traffic congestion \cite{ref9,ref13,ref14}. Numerous approaches based on ITS have been proposed to mitigate traffic congestion and phantom jam. Platoon model and its variations are the most popular solutions \cite{ref15,ref16,ref17,ref18}. In platoon model, a leading vehicle controls all of the following vehicles through wireless communication, makes each vehicle in the string to stay at a desired relative position, thus avoid instability of the vehicle platoon. Leading vehicle tries to bind following vehicles together like locomotive to from a "train" by either uses a preset desired velocity or transmits real-time control information (e.g. position, velocity, etc.) to all the vehicles in the platoon \cite{ref14,ref19}. Inspired by mechanical mass/spring/damper model, bilateral control model was first introduced in \cite{ref20}. In this model, each vehicle in the string adjust its acceleration based on the state information including the distance to and relative velocity of both the vehicle ahead and following, then the equilibrium state of the whole traffic flow could be reached. While within traditional vehicle-following model, each vehicle adjusts its acceleration only with the state information of the preceding vehicle, the instabilities of the traffic flow could be amplified, then phantom jam appears. Damped wave equation was developed in \cite{ref21} and the solutions of the equation showed that damped waves traveling in both directions in the sequence of vehicles under bilateral control, the effect of perturbations were attenuated. \cite{ref22} proved that bilateral control model was chain stable with inserted traffic. Two multi-node version of bilateral control models were discussed in \cite{ref19} which would reach equilibrium state more rapidly.

Unlike the platoon model, there are no leading vehicle nor control node, even without global communication among vehicles in bilateral model, which means bilateral control model is decentralized, vehicles in the string are relatively independent. Control decisions are made mainly depend on the measurements obtained by sensors equipped on host vehicles. Thus bilateral control model has characteristics of distributed control and low communication cost, makes it convenient to realize and operate efficiently under road circumstances. However, the already-proposed bilateral control models still need to be improved in some aspects. Firstly, measurements from on board sensors are the inputs of the control system to make final decisions. Inaccurate sensor information caused either from interior of the sensor or circumstance interference will introduce uncertainties into the system, thereby influencing the final decisions \cite{ref23}. Moreover, Barooah et.al proved that decentralized vehicle control system would suffer from limitations that position information error would be amplified and disturbance in control signals \cite{ref24}. So alleviate the influence of inaccurate sensor information must be considered. Secondly, researchers have shown that vehicle-to-vehicle communication (V2V) could effectively captured real-time vehicle state and the dynamics of traffic jam to control or mitigate phantom jam \cite{ref25,ref26}. Though traditional bilateral control model can work on individual vehicle without any communication among them, it would be more accurate and robust to adopt V2V communications, especially in multi-node bilateral control models proposed in \cite{ref19}.

In this paper, we provide a predictable bilateral control model with local wireless communication to mitigate traffic flow instabilities. Firstly, uncertainties of vehicle state measurements from on-board sensors are explored, Kalman-based algorithm is introduced to alleviate the influence of inaccurate sensor information including distances to host vehicle  and relative velocity of both the vehicles ahead and following. Secondly, we introduce local V2V communications into our model thus the communication delay plays a vital role in the scheme. The influence of communication and system delay from both the vehicle ahead and following with vehicle motion model is considered, a predictive model is proposed to determine accurate acceleration values of the host vehicle.

The rest of the paper is organized as follows:Section II describe the basic models and theories have been already proposed. Section III is divided into two part. In the first part, the problems of the model mentioned above is summarized, in the second part, the predictable bilateral control model was proposed. The experiments are conducted and discussed in section IV, followed by the conclusions in section V.

\section{Related works} 
Treat each vehicle as a control unit, the control decision of the host vehicle (HV) jointly determined by the related vehicles. The input to the control unit are the state vector of corresponding vehicles (i.e difference between the velocity and position between forward and/or backward vehicles), which affects the state of the HV directly (the acceleration in response to the input). So the control unit can be simplified into a system with inputs (i.e relative velocity and position) and output (acceleration). 
\begin{figure}[tbh]
\centering
\includegraphics[scale=0.18]{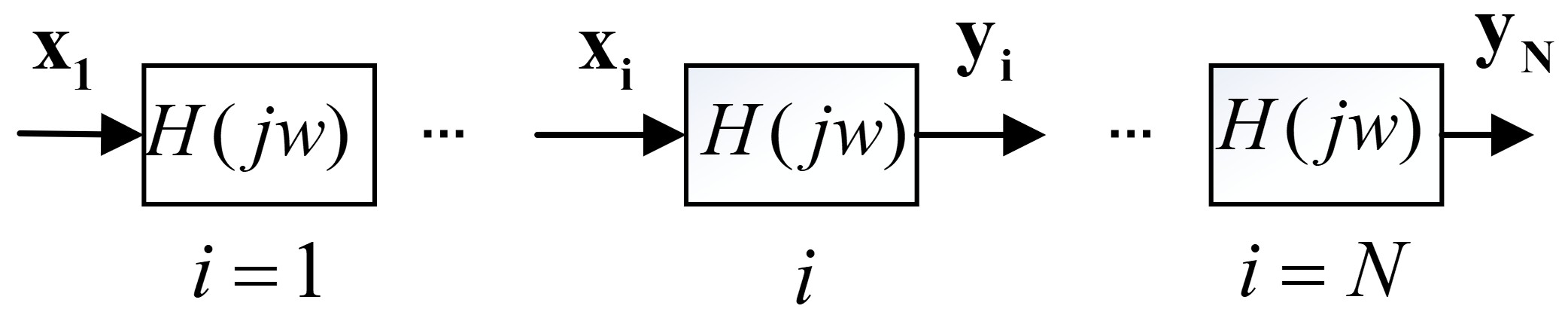}
\caption{Control unit cascade}
\label{pic2_1}
\end{figure}

When $N$ vehicles travel on the straight road, since each vehicle is considered as a control unit, thus $N$ vehicles connected together can be regarded as a consolidated control system as shown in Fig.\ref{pic2_1}. The output of each control unit can be expressed as $\mathbf{y_{i}}=H(j\omega)\mathbf{x_{i}}$, $H(j\omega)$ is the transfer function of the control unit, the total transfer function of the control system is $\mathbf{\frac{y_{n}}{x_{1}}}=\prod_{i=1}^{n} H(j\omega)$.

If the output amplitude of a particular control unit is greater than the input amplitude (i.e $|H(j \omega)|>1$ for some $\omega$), when many control units are cascaded the gain will multiplied together which will lead to larger and larger deviations as more and more control units are cascaded\cite{supressing}.

\subsection{The vehicle following Model (FM)}
\begin{figure}
    \centering
	\includegraphics[scale=0.26]{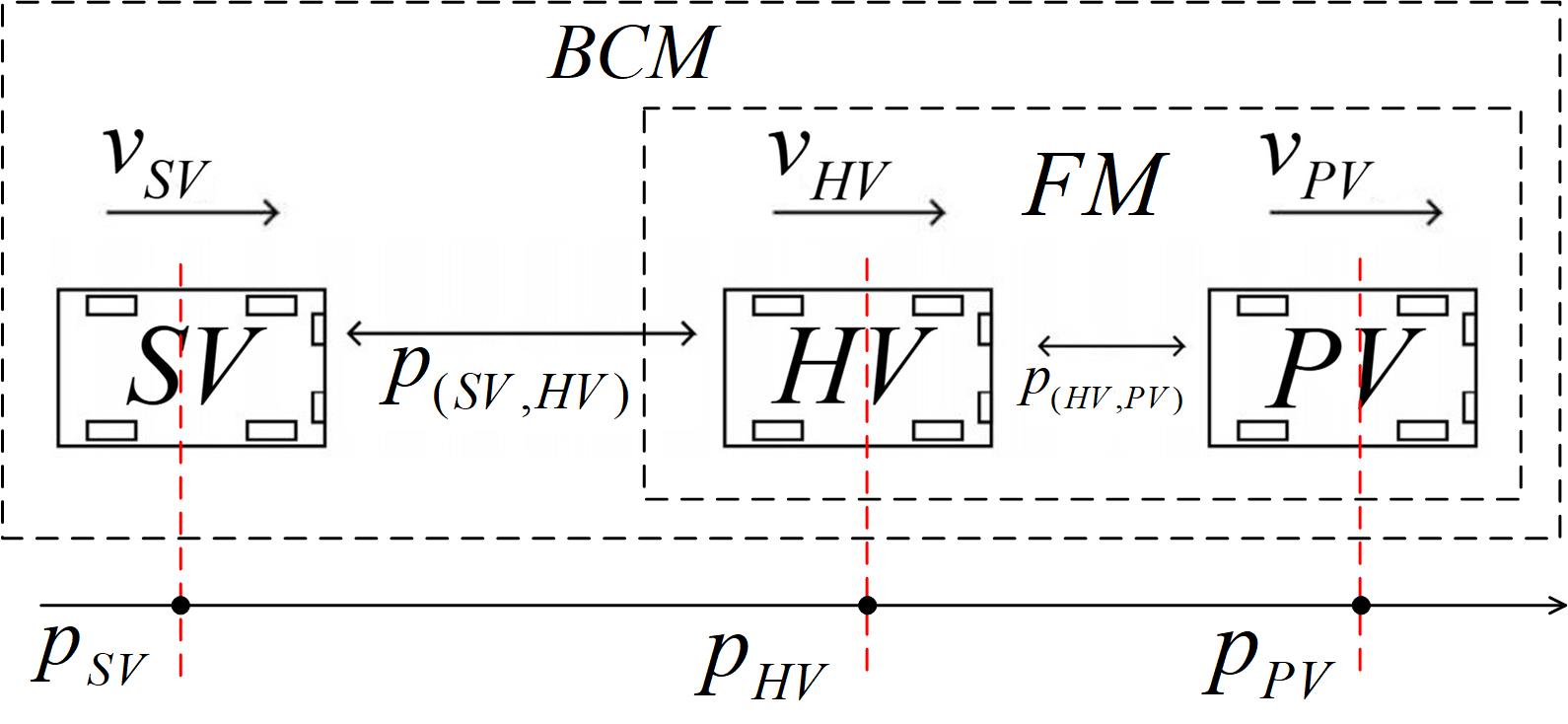}
	\caption{The FM and BCM}
	\label{pic2_2}
    \centering
\end{figure}

The vehicle following model (FM) is shown in small dotted frame in Fig.\ref{pic2_2}, two vehicles travel on a straight road, the preceding vehicle (PV) with the speed of $v_{PV}$ traveling ahead of the HV with the speed of $v_{HV}$, the distance between PV and HV can be marked as $p_{(H V, P V)}=p_{P V}-p_{H V}$ where $p$ represents the position of vehicle.

\begin{figure}
    \centering
	\includegraphics[scale=0.3]{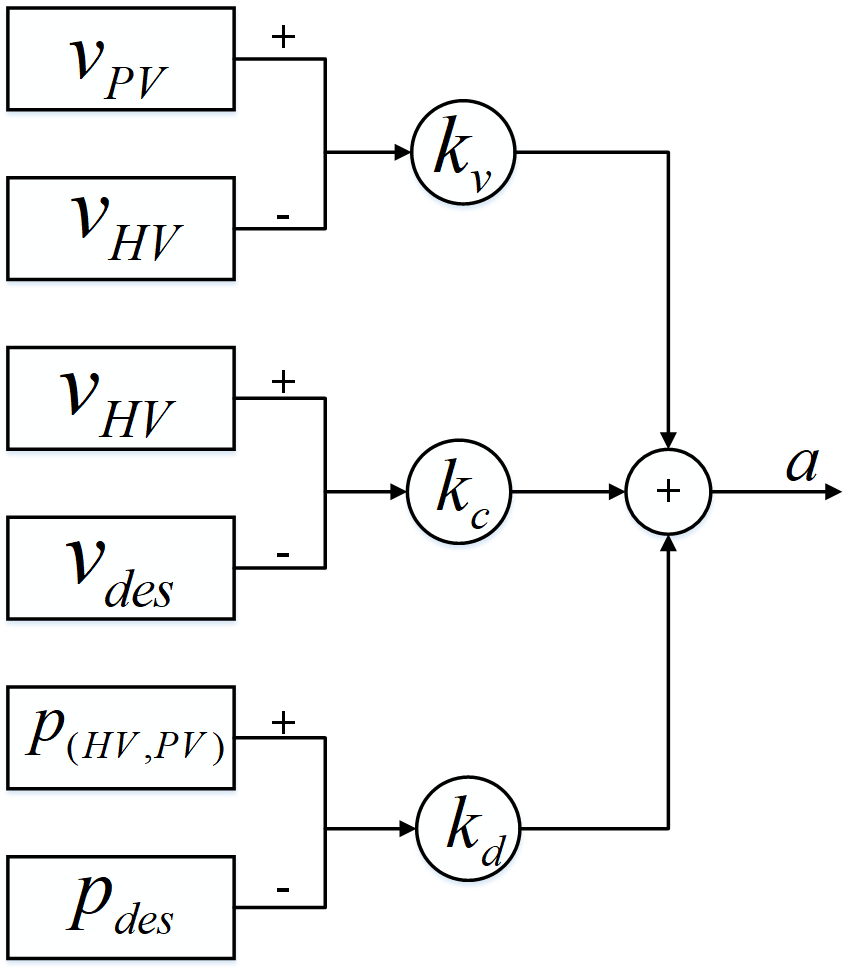}
	\caption{The vehicle following control system }
	\label{pic2_3}
    \centering
\end{figure}

The abstract control model of the FM\cite{supressing} is shown in Fig.\ref{pic2_3}, where $k_{d}$ is the gain of the difference between the real distance ($p_{(H V, P V)}$) and the expected distance ($p_{des}$), $k_{v}$ is the gain of the difference between the two vehicles $v_{(H V, P V)}=v_{P V}-v_{H V}$. $k_{c}$ is the gain of the difference between the expected velocity ($v_{\mathrm{des}}$) and the current velocity ($v_{HV}$).

The acceleration of HV in FM can be calculated as below:
\begin{equation}
\label{eq2_1}
\begin{array}{l}
	a=k_{d}\left(p_{(HV, PV)}-p_{\mathrm{des}}\right)\\
	\quad+k_{v}\left(v_{(H V, P V)}\right)+k_{c}\left(v_{HV}-v_{\mathrm{des}}\right)
\end{array}
\end{equation}
From (\ref{eq2_1}), we can see that HV adjust its acceleration mainly base on the state (i.e relative position $d_{(HV, PV)}$ and relative velocity $v_{(H V, P V)}$) of the PV, which means that the information (the state of PV) only flows in one direction without forward propagation (the HV can acquire the state of succeeding vehicle (SV)).

\subsection{The bilateral control model (BCM)}
Since HV makes driving decision only with the state of PV and itself within FM, while disturbance occurs, the cascade of control unit may amplify the effect, thus congestion may happen, in order to reduce traffic congestion, the bilateral control model (BCM) based on the feedback system was proposed\cite{supressing}.

The BCM is shown in big dotted frame in Fig.\ref{pic2_2}. Three vehicles travel (PV,HV,SV) on a straight road, unlike FM, the SV is also taken into consideration, thus information will flow in both direction (i.e from PV to HV and from SV to HV).

\begin{figure}
	\centering
	\includegraphics[scale=0.3]{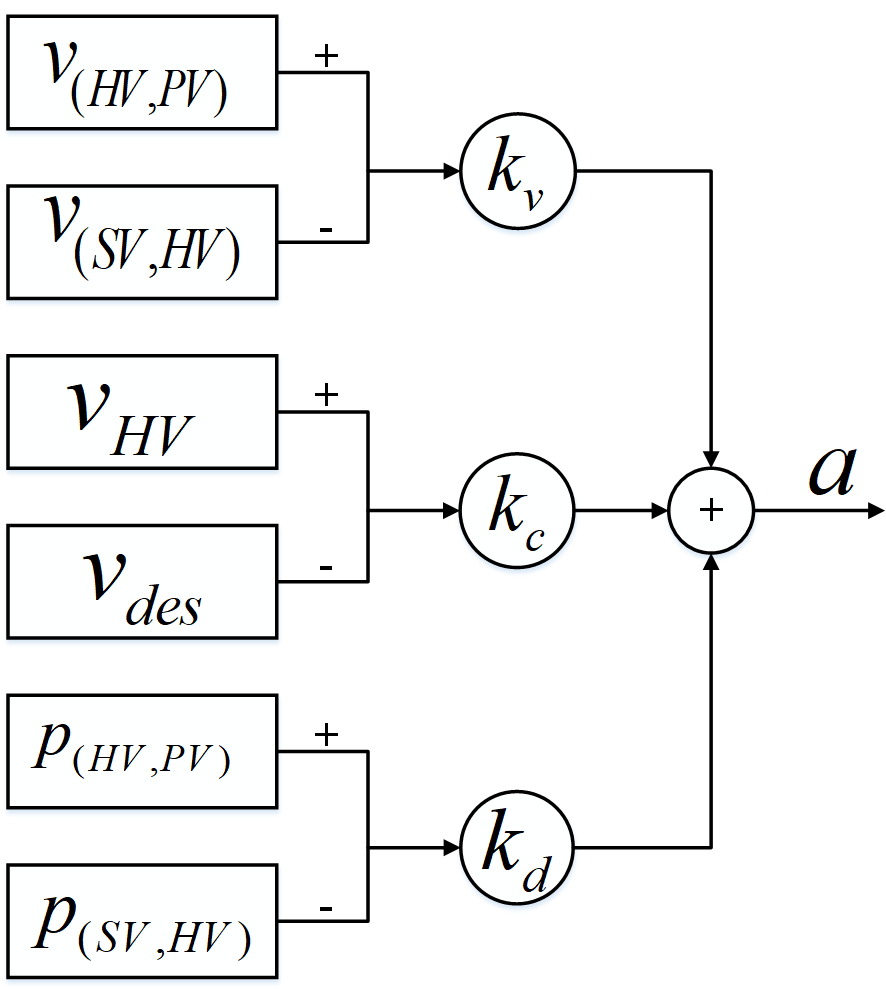}
	\caption{The bilateral control system}
	\label{pic2_5}
	\centering
\end{figure}

Fig.\ref{pic2_5} depicts the abstract control model of the BCM, the acceleration of the HV is determined by the relate state (e.g position, velocity) of the PV and SV. A negative feedback system is formed to adjust the acceleration of the HV.

The acceleration of the HV in BCM can be calculated as below:
\begin{equation}
\label{eq2_2}
\begin{array}{l}
	a_{H V}=k_{d}\left(p_{(H V, P V)}-p_{(S V, H V)}\right)\\
	\quad\quad+k_{v}\left(v_{(H V, P V)}-v_{(S V, H V)}\right)+k_{c}\left(v_{H V}-v_{d e s}\right)
\end{array}
\end{equation}

From (\ref{eq2_2}), it can be known that BCM uses state information of PV and SV. An improved version called multi-node bilateral control model (MBCM) was proposed\cite{multinode}. In this mode of control, when calculating the control decision ($a$), each vehicle will take more vehicles into consideration and put different weight to each vehicle's information. It has been shown that the least squares approach generates sets of coefficients that can damp out low-frequency components of perturbations faster. This means that vehicle system under MBCM will approach an equilibrium state more rapidly than under the traditional version of BCM\cite{multinode}. In this paper we will also take the performance of MBCM into comparison.

\subsection{The Kalman model}
Due to the existence of measurement noise, the measurements are not totally reliable, Kalman model is widely applied in time series analysis to get more accurate system status. The Kalman model is an algorithm that uses a series of measurements observed over time to estimate a jointly probability distribution for each time interval and produces estimates of system status that tend to be more accurate than those based on a single noise-containing measurement alone.

Kalman model basically contain two phase: prediction and update\cite{basekalman}. In prediction phase, the model uses the estimation result of the previous state to make an prediction of the current state. In update phase, the model optimizes the predicted value obtained in the prediction phase by using the measurement of the current time, thus a more accurate state can be estimated.

Basic Kalman model is limited to linear assumption, when the system is linear the basic Kalman can work effectively, but for more complex systems, however, could be nonlinear, so some improved methods such as extended Kalman model (EKF) and unscented Kalman model (UKF) were proposed to resolve this. The EKF uses first-order Taylor expansion to transfer the nonlinear problem into linear problem, but the transformation usually causes the loss of precision and this procedure involves the calculation of Jacobi matrices which is time-consuming\cite{ekfdis,ekfdis2}. The UKF uses a deterministic sampling technique known as the unscented transformation (UT) to pick a minimal set of sample points (called sigma points) around the mean,  thus the UKF can process multi-sensor data efficiently, but the calculation process of UKF is similar to second-order Taylor's expansion which is also time-consuming and complex\cite{ukf}\cite{ukfmulti}.

Since our system only use the linear parameter like position and velocity of single target and the system require real-time data process, so the basic Kalman model is adopted in this paper.

\section{The predictable bilateral control model}

\subsection{Problem descriptions}
The models we mentioned above (i.e FM, BCM, MBCM) are all in idea situation, which means they haven't consider the influence of instability factors (e.g. errors, delay). The vehicle state information is mainly acquired by the sensors which will have measurement error inevitably, in addition, HV need to sense the information of PV and SV in time which require low time delay. Thus the instabilities in an vehicles queue basically come from two aspects: errors and delay.

\subsubsection{Errors}
The information acquired by the sensors is affected by internal and external noise, both will cause errors in our system\cite{itsxiang}.

The first is internal noise from the radar sensor, taking the FMCW radar sR-1200e $24GHz$ for example\cite{fmcwbasic}, It's minimum accuracy is 0.6$m$, which means this FMCW radar has an error of 0.6$m$, in addition, the inner noise generated by the inter electronic components and phase noise also will result in measurement error\cite{fmcwnoise}\cite{itsxiang}.

The second is external noise, when vehicles move on the road, the environmental noise, like the raining or snowing day, will reduce the signal-to-noise ratio (SNR) and have a uncertain impact on measurement process.

After considering both the internal noise and the external noise, the measurement error is about 2\%-10\% for most vehicle system.
\subsubsection{Delay}
There are two main type of delay in vehicle system: communication delay and system delay.

In the bilateral control system, information is transmitted in both directions. In order to transmit the information through the system, the communication delay should take into consideration. The communication of intelligent transportation systems generally follows the V2X standard and most of the V2X support the V2V communication. The mostly used V2X communication methods are show in Table.\ref{tab1}, it can be seen that for most V2X, the delay is 10$ms$-80$ms$ and we mark it as $t_{comm}$.
\begin{table*}
	\centering
	\caption{The common V2X method}
	\label{tab1}
	\begin{tabular}{|c|c|c|c|c|}
		\hline
		\multirow{2}{*}{Characteristic} & \multicolumn{4}{|c|}{Technical} \\
		\cline{2-5}
		& IEEE802.11p & IEEE802.11px & LTE-A PRO & 5G mmWave \\
		\hline
		Frequency Band & 5.85-5.925 $GHz$ & 5.85-5.925 $GHz$ & 5.72-5.75 $GHz$ & 57.05-64 $GHz$ \\
		\hline
		Bandwidth & 10 $MHz$ & 10 $MHz$ & $ \ge $640 $MHz$ & 2.16 $GHz$ \\
		\hline
		Range & 1$KM$ & 1$KM$ & 30$KM$ & 50$M$ \\
		\hline
		Bit rate & $3-27Mbps$ & $ \ge $60 $Mbps$ & $ \ge $3 $Gbps$ & $ \ge $7 $Gbps$ \\
		\hline	
		Delay & 10 $ms$ & 10 $ms$ & 20-80 $ms$ & 10 $ms$ \\
		\hline	
		Broadcast & Y & Y & Y & N \\
		\hline	
		V2V Support & Y & Y & Y & Y \\
		\hline
		MIMO & Y & Y & Y & Y \\
		\hline	
	\end{tabular}
\end{table*}

In order to ensure safety and reliability, the ITS is required as a real-time system to ensure that tasks (i.e compute process, system barking) can completed within a limited time. Recently years, companies such as Google, Baidu's Apollo, and Cruise are all conducting related researches. Due to the need of perceive and calculate fairly large of information in real time, the response time of most ITS at present is limited from 50$ms$ to 120$ms$\cite{delaysupport} and we mark it as $t_{sys}$.

Take both the system delay and communication delay into consideration, the total delay of a system is as below:
$$t_{total}=t_{comm}+t_{sys}$$

In our proposed model, the calculation will be performed for each time interval $\Delta{t}$ and the total delay is limited from 60$ms$ to 200$ms$, in order to make the calculation perform normally, so the time interval should meet: $\Delta{t} \geq 60ms$.

In summary, when apply the model to the real scenarios, errors and delay are the mainly two instability factors need to take into consideration, thus in the our proposed model, we will focus on reduce the impact cause by the instabilities.

\subsection{System overview}
\begin{figure*}
	\centering
	\includegraphics[scale=0.3]{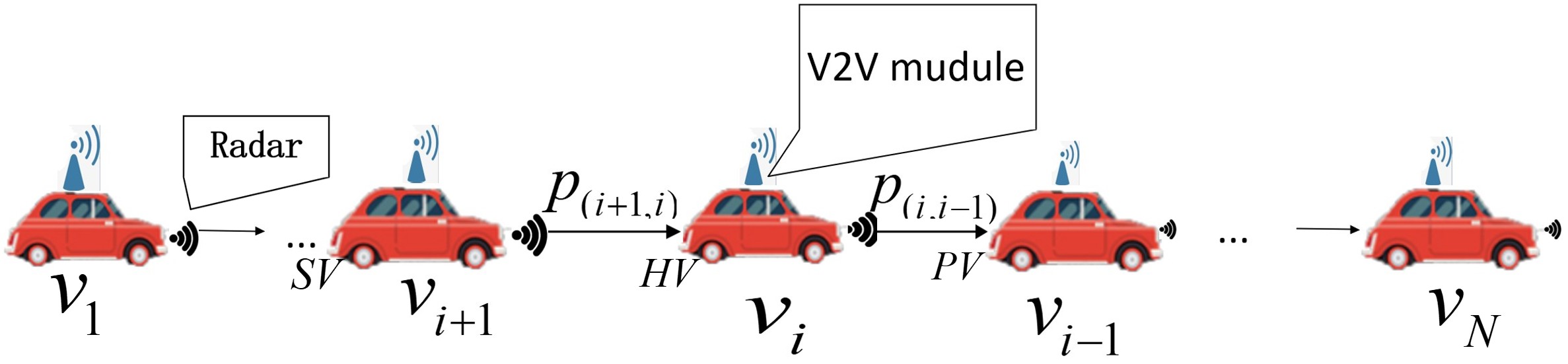}
	\caption{Vehicles in traffic flow}
	\centering
	\label{pic4_1}
\end{figure*}

Assuming that all vehicles are on a one-way straight road as shown in Fig.\ref{pic4_1}. There are  $N$ vehicles, numbered $1, 2,.i..N$, except for the first vehicle $1$ and last vehicle $N$, each vehicle has preceding vehicle (PV) and succeeding vehicle (SV). All vehicles are heading in one direction. At the beginning, the vehicles evenly spaced on the road with the distance $d_0$ and the velocity of each vehicle is $v_0$. 

The state of a vehicle at time $t$ mainly consist of position $p_t$ and velocity $v_t$ and the control decision of a vehicle is the acceleration $a$. So the related state vectors are defined in Tab.\ref{tabps}.
\begin{table}[H]
	\centering
	\caption{Related parameters}
	\label{tabps}
	\begin{tabular}{@{}ccll@{}}
		\toprule
		State      & ${\mathbf{x}}_{t}=\left[p_{t}, v_{t}\right]^{T}$  & Measurement & \multicolumn{1}{c}{$\bar{\mathbf{x}_{t}}=[\bar{p}_{t}, \bar{v}_{t}]^{T}$} \\ \midrule
		Prediction & $\hat{\mathbf{x}_{t}}=[\hat{p}_{t}, \hat{v}_{t}]^{T}$ & Estimation   & $\tilde{\mathbf{x}_{t}}=[\tilde{p}_{t}, \tilde{v}_{t}]^{T}$           \\ \bottomrule
	\end{tabular}
\end{table}

Take arbitrary vehicle $i$ in the traffic flow for example, we mark it as HV, thus the index of PV and SV is $i-1$ and $i+1$ respectively, the state of PV relative to HV can be defined as follow:
\begin{equation}
	\begin{array}{l}
		p_{(i, i-1)}=p_{i-1}-p_{i} \\
		v_{(i, i-1)}=v_{i-1}-v_{i} \\
	\end{array}
	\label{define2}
\end{equation}

The acceleration of PV relative to HV can be defined similarly:
$$
a_{(i, i-1)}=a_{i-1}-a_{i}
$$

And the relative state of SV relative to HV can be defined in the similar way as shown above.

HV has two modules, one is the vehicle radar (e.g FMCW radar) and the other is the V2V module. At time $t$, with the radar, HV can acquire the information of PV (i.e $\bar{p}_{(i, i-1),t}$ and $\bar{v}_{(i,i-1),t}$) and with the V2V module, HV can receive the information of SV (i.e $p_{(i+1, i),t}$ and $v_{(i+1,i),t}$) as shown in Fig.\ref{pic4_1}.
\begin{figure*}
	\centering
	\includegraphics[scale=0.35]{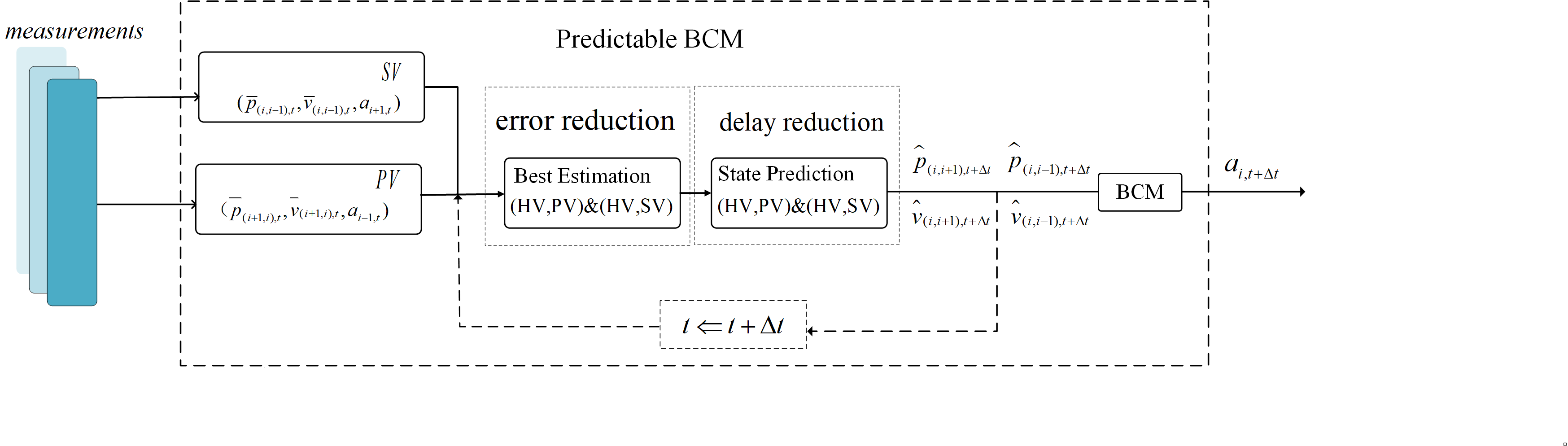}
	\caption{The predictable bilateral control model}
	\centering
	\label{overview}
\end{figure*}

The inputs of PBCM is the measurements from the radar and V2V module, and as shown in Fig.\ref{overview}, PBCM mainly consists of two procedures: best estimation and state prediction.
\begin{itemize}
	\item \textbf{Best Estimation}: Combine the measurements and prediction predicted at last time ($t-\Delta{t}$), the best estimation can be acquired which means the errors would be reduced.
	\item \textbf{State prediction}: With the best estimation, HV can make the state prediction of related vehicles ( SV, HV, PV) for next time ($t+\Delta{t}$), which means the vehicles state of time $t+\Delta{t}$ is predicted at time $t$.
\end{itemize}

With the procedures above, the influence of errors and delay could be eliminated, and the state of vehicles could be predicted, thus the control decision of HV can be made in advance with more accurate vehicle information and low time delay. 

\subsection{The predictable bilateral control model (PBCM)}
In order to reduce the influence of instabilities (i.e errors, delay) when apply BCM into real scenario, in this section a predictable bilateral control model (PBCM) with local wireless communication and Kalman model to mitigate traffic flow instability is proposed.

In the queue, the last vehicle don't have the SV, but it has the PV which is consistent to the FM, and we set the first vehicle in the queue travels with a constant velocity. Thus the control decision ($a_{i, t}$) of vehicle $i$ at time $t$ under idea situation can be described as below:
\begin{equation}
	a_{i, t}=
	\left\{\begin{aligned}
		\begin{array}{lr}
		0 \quad\quad\quad\quad\quad\quad\quad\quad\quad\quad\quad\quad\quad\quad\quad i=1 \\
		k_{d}\left(p_{(i, i-1), t}-p_{(i+1, i), t}\right) \\
		\quad+k_{v}\left(v_{(i, i-1), t}-v_{(i+1, i), t}\right) \\
		\quad+k_{c}\left(v_{i, t}-v_{d e s}\right) \quad\quad i=2,3, \ldots, N-1 \\
		k_{d}\left(p_{(i, i-1), t}-p_{d e s}\right) \\
		\quad+k_{v}\left(v_{(i, i-1), t}\right)+k_{c}\left(v_{i, t}-v_{d e s}\right) \quad i=N
	\end{array}
	\end{aligned}\right.
\label{eq3_2}
\end{equation}

\begin{figure}[h]
	\includegraphics[scale=0.25]{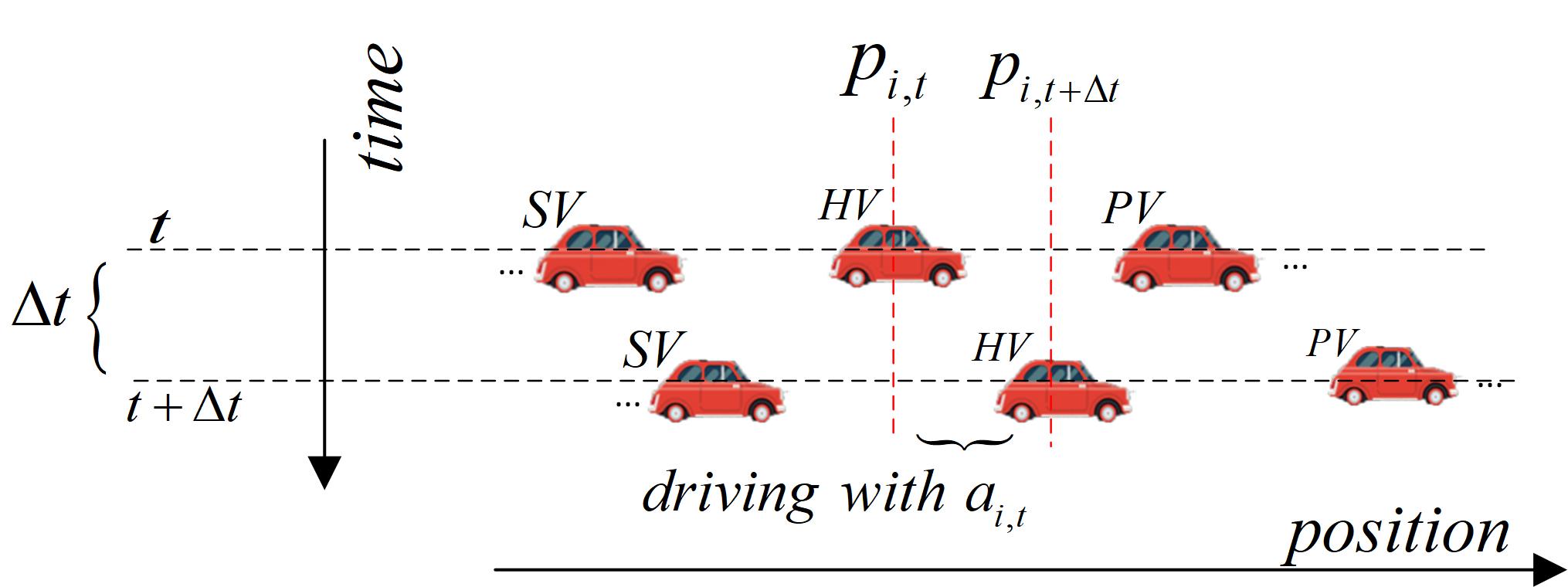}
	\caption{related state of a single bilateral model}
	\label{pic_25_1}
\end{figure}

In the idea situation, the HV can get the relative velocity and position of SV and PV exactly at time $t$, then put the relative state into equation.\ref{eq3_2} the control decision($a_{i,t}$) can be get, so between $t$ and $t+\Delta t$ the vehicle will travel with acceleration $a_{i,t}$. The procedure can be shown in Fig.\ref{pic_25_1}

When we apply the BCM into real scenario, from the problem descriptions, we know that the errors and delay are inevitable. Firstly, from the (\ref{eq3_2}), it can be seen that the calculation of $a_{i,t}$ mainly rely on the related position $p$ and velocity $v$, so the more accurate the $p$ and $v$ is, the more better driving decision we can make, but due to errors, it is impossible for HV to obtain accurate state of related vehicles. Secondly, as shown in Fig.\ref{pic_25_1}, under the ideal situation, we assume that when the control decision ($a_{i,t}$) calculated, HV will drive with $a_{i,t}$ immediately, but in real scenario it is impossible for HV to reach $a_{i,t}$ at time $t$, the acceleration we calculated will take effect at $t+\Delta t$, which means the system has a delay $\Delta t$.

From the relative state has been defined in (\ref{define2}), the state vector of PV relative to HV at time $t$ can be written as $\mathbf{x}_{(i, i-1), t}=\left[p_{(i, i-1), t}, v_{(i, i-1), t}\right]^{T}$, the state vector of HV relative to SV at time $t$ can be written as $\mathbf{x}_{(i+1, i), t}=\left[p_{(i+1, i), t}, v_{(i+1, i), t}\right]^{T}$.

The PBCM mainly contain two procedure: state prediction and best estimation.

\subsubsection{state prediction}
As show in Fig.\ref{pic_25_1}, the current time is $t$, the next time $t+{\Delta t}$ is marked as $t_{1}$, in order to explain the all procedure in detail, take arbitrary vehicle $i$ (expect for the first and last) in the traffic flow for example, so the related vehicles are $i-1$ (PV) and $i+1$ (SV). At time $t$, HV can acquire the relative measurement value of PV: $\bar{\mathbf{x}}_{(i, i-1), t}=\left[\bar{p}_{(i, i-1), t}, \bar{v}_{(i, i-1), t}\right]^{T}$ and SV: $\bar{\mathbf{x}}_{(i+1, i), t}=\left[\bar{p}_{(i+1, i), t}, \bar{v}_{(i+1, i), t}\right]^{T}$, and with the relative predicted value of PV: $\hat{\mathbf{x}}_{(i, i-1), t}=\left[\hat{p}_{(i, i-1), t}, \hat{v}_{(i, i-1), t}\right]^{T}$ and SV: $\hat{\mathbf{x}}_{(i+1, i), t}=\left[\hat{p}_{(i+1, i), t}, \hat{v}_{(i+1, i), t}\right]^{T}$ the best estimation of SV: $\tilde{\mathbf{x}}_{(i+1, i), t}=\left[\tilde{p}_{(i+1, i), t}, \tilde{v}_{(i+1, i), t}\right]^{T}$ and PV: $\tilde{\mathbf{x}}_{(i, i-1), t}=\left[\tilde{p}_{(i, i-1), t}, \tilde{v}_{(i, i-1), t}\right]^{T}$ can be acquired. Once the best estimation acquired, the relative state of HV and PV at next time $t_{1}$ can be predicted with (\ref{eq03_01})
\begin{equation}
\left\{\begin{array}{l}
	\hat{p}_{(i,i-1), t_{1}}=\tilde{p}_{(i,i-1), t}+\tilde{v}_{(i,i-1), t} \times \Delta t+{a}_{(i,i-1), t} \times \frac{\Delta t^{2}}{2} \\
	\hat{v}_{(i,i-1), t_{1}}=\tilde{v}_{(i,i-1), t}+{a}_{(i,i-1), t} \times \Delta t
\end{array}\right.
\label{eq03_01}
\end{equation}
Similarly, the relative state of HV and SV at next time $t_{1}$ can be predicted with Eq.\ref{eq03_02}
\begin{equation}
\left\{\begin{array}{l}
	\hat{p}_{(i+1,i), t_{1}}=\tilde{p}_{(i+1,i), t}+\tilde{v}_{(i+1,i), t} \times \Delta t+{a}_{(i+1,i), t} \times \frac{\Delta t^{2}}{2} \\
	\hat{v}_{(i+1,i), t_{1}}=\tilde{v}_{(i+1,i), t}+{a}_{(i+1,i), t} \times \Delta t
\end{array}\right.
\label{eq03_02}
\end{equation}

Then we get:
\begin{equation}
\left[\begin{array}{l}
	\hat{p}_{(i, i-1), t_{1}} \\
	\hat{v}_{(i, i-1), t_{1}}
\end{array}\right]=F\left[\begin{array}{l}
	\tilde{p}_{(i, i-1), t} \\
	\tilde{v}_{(i, i-1), t}
\end{array}\right]+B a_{(t, t-1), t}
\label{eq03_03}
\end{equation}
\begin{equation}
\left[\begin{array}{l}
	\hat{p}_{(i+1,i), t_{1}} \\
	\hat{v}_{(i+1,i), t_{1}}
\end{array}\right]=F\left[\begin{array}{l}
	\tilde{p}_{(i+1, l), t} \\
	\tilde{v}_{(i+1, i), t}
\end{array}\right]+B a_{(i+1, i), t}
\label{eq03_04}
\end{equation}

We define:
$$
\begin{array}{l}
	F=\left[\begin{array}{ll}
		1 & \Delta t \\
		0 & 1
	\end{array}\right] \quad , \quad
	B=\left[\begin{array}{l}
		\frac{\Delta t^{2}}{2} \\
		\Delta t
	\end{array}\right]
\end{array}
$$

${F}$ is the state transition matrix, represents the state transition relationship of vehicle from ${t}$ to ${t_1}$. ${B}$ is the control matrix, it represents the effect of acceleration on the change of acceleration to vehicle state. The state prediction procedure can be written as below:
\begin{equation}
	\left\{\begin{array}{l}
		\hat{x}_{(i,i-1), t_{1}}=F \tilde{x}_{(i,i-1), t}+B {a}_{(i,i-1), t} \\
		\hat{x}_{(i+1,i), t_{1}}=F \tilde{x}_{(i+1,i), t}+B {a}_{(i+1,i), t}
	\end{array}\right.
\label{eq03_06}
\end{equation}
Apply the predicted parameter ${\hat p_{(i,i-1),{t_1}}}$, ${\hat p_{(i+1,i),{t_1}}}$ and ${\hat v_{(i,i-1),{t_1}}}$, ${\hat v_{k+1,{t_1}}}$ to calculate $a$, Then the $a$ at time ${t_1}$ can be acquired at time ${t}$ from Eq.\ref{eq03_08}, which could eliminate the influence of Delay.
\begin{equation}
	a_{i, t_{1}}=
	\left\{\begin{aligned}
		\begin{array}{lr}
			0 \quad\quad\quad\quad\quad\quad\quad\quad\quad\quad\quad\quad i=1 \\
			k_{d}\left(\hat{p}_{(i,i-1), t_{1}}-\hat{p}_{(i+1,i), t_{1}}\right) \\
			\quad+k_{v}\left(\hat{v}_{(i,i-1), t_{1}}-\hat{v}_{(i+1,i), t_{1}}\right) \\
			\quad+k_{c}\left(\hat{v}_{i, t_{1}}-v_{d e s}\right) \quad i=2,3, \ldots, N-1 \\
			k_{d}\left(\hat{p}_{(i, i-1), t_{1}}-p_{d e s}\right) \\
			+k_{v}\left(\hat{v}_{(i, i-1), t_{1}}\right)+k_{c}\left(\hat{v}_{i, t_{1}}-v_{d e s}\right) i=N
		\end{array}
	\end{aligned}\right.
	\label{eq03_08}
\end{equation}

So far, the state prediction procedure is accomplish, and the control decision is acquired for time $t_{1}$ at time $t$, so when come to time $t_{1}$,  the acceleration can take effect immediately with no delay.

\subsubsection{Best estimation}
Take the errors into consideration,  the matrix $Q{_{}}$ is used to represent the instability.
\begin{equation}
P_{{{\rm{t}}_1}}^{\prime} = F{P_{{t}}}{F^T} + Q
\label{eq03_07}
\end{equation}

In (\ref{eq03_07}) show the update  of the prior state covariance matrix $P_{{{\rm{t}}_1}}^{\prime}$ of next time, the $Q$ influencing the priori status update procedure, which represent transfer relationship of environmental noise in the real scenario between ${t}$ and ${t_{1}}$

Then come to time ${t_1}$, similar to Kalman model, the relative state of PV:$\bar{x}_{(i, i-1), t_{1}}$ and SV:$\bar{x}_{(i+1, i), t_{1}}$ can be acquired, with the prediction on last time as show in Eq.\ref{eq03_01}, the best estimation can be calculated with (\ref{eq03_09})
\begin{equation}
	\begin{array}{l}
		\tilde{x}_{(i,i-1), t_{1}}=\hat{x}_{(i,i-1), t_{1}}+K_{t_{1}}\left(\bar{x}_{(i,i-1), t_{1}}-H \hat{x}_{(i,i-1), t_{1}}\right) \\
		\tilde{x}_{(i+1,i), t_{1}}=\hat{x}_{(i+1,i), t_{1}}+K_{t_{1}}\left(\bar{x}_{(i+1,i), t_{1}}-H \hat{x}_{(i+1,i), t_{1}}\right)
	\end{array}
\label{eq03_09}
\end{equation}

$H$ is the observation matrix and ${K_{{t_1}}}$ is Kalman coefficient, which is the key parameter of Kalman model. It's main function is to merge the observation procedure and the prediction procedure to obtain the state more close to the real state, and the ${K_{{t_1}}}$ can be calculated from equation Eq.\ref{eq03_10} and the estimate covariance can be update with (\ref{eq03_11})
\begin{equation}
	\label{eq03_10}
	{K_{{t_1}}} = P_{{t_1}}^{\prime} {H^T}{\left( {HP_{{t_1}}^{\prime} {H^T} + R} \right)^{ - 1}}
\end{equation}
\begin{equation}
	{P_{{t_1}}} = \left( {I - {K_{{t_1}}}H} \right)P_{{t_1}}^{\prime}
\label{eq03_11}
\end{equation}

Similar to Eq.\ref{eq03_01}, at time ${t_1}$, the next state ${\hat x_{(i,i-1),{t_{\rm{2}}}}}$ can be predicted as below:
\begin{equation}
\left\{\begin{array}{l}
	\hat{p}_{(i,i-1), t_{2}}=\tilde{p}_{(i,i-1), t_{1}}+\tilde{v}_{(i,i-1), t_{1}} \times \Delta t+{a}_{(i,i-1), t_{1}} \times \frac{\Delta t^{2}}{2} \\
	\hat{v}_{(i,i-1), t_{2}}=\tilde{v}_{(i,i-1)}+{a}_{(i,i-1), t_{1}} \times \Delta t
\end{array}\right.
\label{eq03_12}
\end{equation}
And the prediction of relative state SV and HV is the same as show in (\ref{eq03_12}). 

The Algorithm of PBCM is as follow:
\begin{algorithm}[H]
	\caption{\quad\quad\quad\quad PBCM}
	\hspace*{0.02in} {\bf Initialization:} 
	$t=0$, $Q$, $H$, $P_{0}$, $F$, $B$, $K_{0}$\\
	\hspace*{0.02in} {\bf Output:}
	$a_{(i,t)}$
	\begin{algorithmic}
		\State Radar Input : $\bar{x}_{(i-1, i), t}$
		\State V2V Input : $\bar{x}_{(i+1, i), t}$
		\State Prediction at last time: $\hat{x}_{(i,i-1), t}$,$\hat{x}_{(i+1, i), t}$
		
		\State Best estimation\\
		\qquad $\tilde{x}_{(i,i-1),t}=\hat{x}_{(i, i-1), t}+K_{t}\left(\bar{x}_{(i, i-1), t}-H \hat{x}_{(i, i-1), t}\right)$\\
		\qquad $\tilde{x}_{(i+1,i), t}=\hat{x}_{(i+1, i), t}+K_{t}\left(\bar{x}_{(i+1, i), t}-H \hat{x}_{(i+1, i), t}\right)$
		\State State prediction\\
		\qquad $\hat{x}_{(i, i-1), t_{1}}=F \tilde{x}_{(i, i-1), t}+B a_{(i, i-1), t}$\\
		\qquad $\hat{x}_{(i+1, i), t_{1}}=F \tilde{x}_{(i+1, i), t}+B a_{(i+1, i), t}$\\
		\qquad  Compute the $a_{(i,t_{1})}$ for next time with BCM

		\State Update for time $t_1$\\
		\qquad $P_{\mathrm{t}_{1}}^{\prime}=F P_{t} F^{T}+Q$\\
		\qquad $K_{t_{1}}=P_{t_{1}}^{\prime} H^{T}\left(H P_{t_{1}}^{\prime} H^{T}+R\right)^{-1}$ \\
		\qquad $P_{t_{1}}=\left(I-K_{t_{1}} H\right) P_{t_{1}}^{\prime}$
	\end{algorithmic}
\end{algorithm}

With the state prediction procedure to eliminate the influence of delay and the best estimation to reduce errors, the PBCM is proposed. In the next section, a simulation platform will be built which is similar to the real scenario and  a series of experiments will be simulated to show the performance of PBCM. 

\section{Results and Analysis}
\subsection{Simulation platform}
In our simulation, we developed a virtual one-way straight road with 40 vehicles placed on it, simulation parameters are summarized in Tab.\ref{tab4}. All vehicles are placed on the road evenly with the distance of 37.5m and speed of 30$m/s$ at the beginning of the simulation. Consider the comfort of driver and passengers, the driving decision $a_{i, t}$ would follow the rule\cite{itsxiang}:
$$
a_{i, t}=\left\{\begin{array}{cc}
	3 m / s^{2} & a_{i, t} \geq 3 m / s^{2} \\
	a_{i, t} & -3 m / s^{2}<a_{i, t}<3 m / s^{2} \\
	-3 m / s^{2} & a_{i, t} \leq-3 m / s^{2}
\end{array}\right.
$$
\begin{table}[H]
	\centering
	\caption{Parameters in simulation}
	\label{tab4}
	\begin{tabular}{|l|l|}
		\hline
		Name		         & Value										  \\ \hline
		Road length	$l_1$	     & 1500$m$										  \\ \hline
		Road width	$l_2$	     & 5$m$											  \\ \hline
		Total duration		 & 500s											  \\ \hline
		Initial Distance $d_0$ & 37.5$m$                         \\ \hline
		Vehicle length  $l_3$     & L=5m                                           \\ \hline
		Number of vehicles $N$  & 40                                   \\ \hline
		Max/Min Acceleration & $\mathrm{a}=\pm 3 \mathrm{m} / \mathrm{s}^{2}$ \\ \hline
		Initial velocity $v_0$   & 30$m/s$										  \\ \hline
		Braking time         & $t=2 s$                                        \\ \hline
	\end{tabular}
\end{table}
In our simulation, we used four different models (i.e FM, BCM, MBCM, PBCM) to control the vehicles motion, the corresponding simulation cases are summarized in Tab.\ref{tab5}. 

The running time of the whole simulation system is 500$s$, from section II, we already know that the system will be unstable and amplify the instability in FM. All simulation cases would run with the FM for the first 10$s$, after that, four different models as shown in the table would run from $10-500s$. In traffic flow, if arbitrary one of vehicle brakes suddenly (i.e disturbance occurs), a chain reaction will appear, which make the distance between vehicles unbalanced thus causing a certain degree of congestion in the system\cite{supressing}. In our simulations, the vehicle in the traffic flow (index 5) will brake with acceleration $-3 / \mathrm{m}^{2}$, and the braking process lasts for 2$s$ at 10$s$. In addition, combination of delay and error settings will have different effects on the operation of the vehicle, as we have mentioned the delay is 70$ms$-200$ms$ and the error is 2\%-10\%, so in order to compare different models more comprehensive, the scenario we designed will have different error (i.e 3\% and 10\%) and delay (i.e 100$ms$ and 200$ms$) as shown in Tab.\ref{tab5}

In the following subsection, we show the performances of error reduction and compared the performance of different models.
\begin{table*}[]
	\caption{Cases}
	\label{tab5}
	\centering
	\begin{tabular}{|c|c|cc|l|l|}
		\hline
		0-10(s)                                                                              & \multicolumn{5}{c|}{10(s)-100(s)}                                                                                                                                                                                                                                   \\ \hline
		\multirow{17}{*}{\begin{tabular}[c]{@{}c@{}}vehicle following model\cite{supressing} \\ (FM)\end{tabular}} & \multicolumn{1}{l|}{Model}                                                                            & \multicolumn{1}{l|}{Acceleration and braking time}                                                & \multicolumn{1}{l|}{Measurement error} & Delay & CaseId \\ \cline{2-6} 
		& \multirow{4}{*}{\begin{tabular}[c]{@{}c@{}}vehicle following model\cite{supressing}\\ (FM)\end{tabular}}                   & \multicolumn{1}{c|}{\multirow{4}{*}{\begin{tabular}[c]{@{}c@{}}$-3 m / s^{2}$\\ 2s\end{tabular}}} & \multirow{2}{*}{3\%}                   & 0.1s  & 1      \\ \cline{5-6} 
		&                                                                                                       & \multicolumn{1}{c|}{}                                                                             &                                        & 0.2s  & 2      \\ \cline{4-6} 
		&                                                                                                       & \multicolumn{1}{c|}{}                                                                             & \multirow{2}{*}{10\%}                  & 0.1s  & 3      \\ \cline{5-6} 
		&                                                                                                       & \multicolumn{1}{c|}{}                                                                             &                                        & 0.2s  & 4      \\ \cline{2-6} 
		& \multirow{4}{*}{\begin{tabular}[c]{@{}c@{}}Bilateral control model\cite{supressing}\\ (BCM)\end{tabular}}    & \multicolumn{1}{c|}{\multirow{4}{*}{\begin{tabular}[c]{@{}c@{}}$-3 m / s^{2}$\\ 2s\end{tabular}}} & \multirow{2}{*}{3\%}                   & 0.1s  & 5      \\ \cline{5-6} 
		&                                                                                                       & \multicolumn{1}{c|}{}                                                                             &                                        & 0.2s  & 6      \\ \cline{4-6} 
		&                                                                                                       & \multicolumn{1}{c|}{}                                                                             & \multirow{2}{*}{10\%}                  & 0.1s  & 7      \\ \cline{5-6} 
		&                                                                                                       & \multicolumn{1}{c|}{}                                                                             &                                        & 0.2s  & 8      \\ \cline{2-6} 
		& \multirow{4}{*}{\begin{tabular}[c]{@{}c@{}}Multi-node bilateral control model\cite{multinode}\\ (MBCM)\end{tabular}}  & \multicolumn{1}{c|}{\multirow{4}{*}{\begin{tabular}[c]{@{}c@{}}$-3 m / s^{2}$\\ 2s\end{tabular}}} & \multirow{2}{*}{3\%}                   & 0.1s  & 9      \\ \cline{5-6} 
		&                                                                                                       & \multicolumn{1}{c|}{}                                                                             &                                        & 0.2s  & 10     \\ \cline{4-6} 
		&                                                                                                       & \multicolumn{1}{c|}{}                                                                             & \multirow{2}{*}{10\%}                  & 0.1s  & 11     \\ \cline{5-6} 
		&                                                                                                       & \multicolumn{1}{c|}{}                                                                             &                                        & 0.2s  & 12     \\ \cline{2-6} 
		& \multirow{4}{*}{\begin{tabular}[c]{@{}c@{}}Predictable bilateral control model\\ (PBCM)\end{tabular}} & \multicolumn{1}{c|}{\multirow{4}{*}{\begin{tabular}[c]{@{}c@{}}$-3 m / s^{2}$\\ 2s\end{tabular}}} & \multirow{2}{*}{3\%}                   & 0.1s  & 13     \\ \cline{5-6} 
		&                                                                                                       & \multicolumn{1}{c|}{}                                                                             &                                        & 0.2s  & 14     \\ \cline{4-6} 
		&                                                                                                       & \multicolumn{1}{c|}{}                                                                             & \multirow{2}{*}{10\%}                  & 0.1s  & 15     \\ \cline{5-6} 
		&                                                                                                       & \multicolumn{1}{c|}{}                                                                             &                                        & 0.2s  & 16     \\ \hline
	\end{tabular}
\end{table*}

\subsection{Performances of error reduction}
\begin{figure}
	\centering
	\subfigure[Real velocity and velocity in BCM]{
		\includegraphics[scale=0.21]{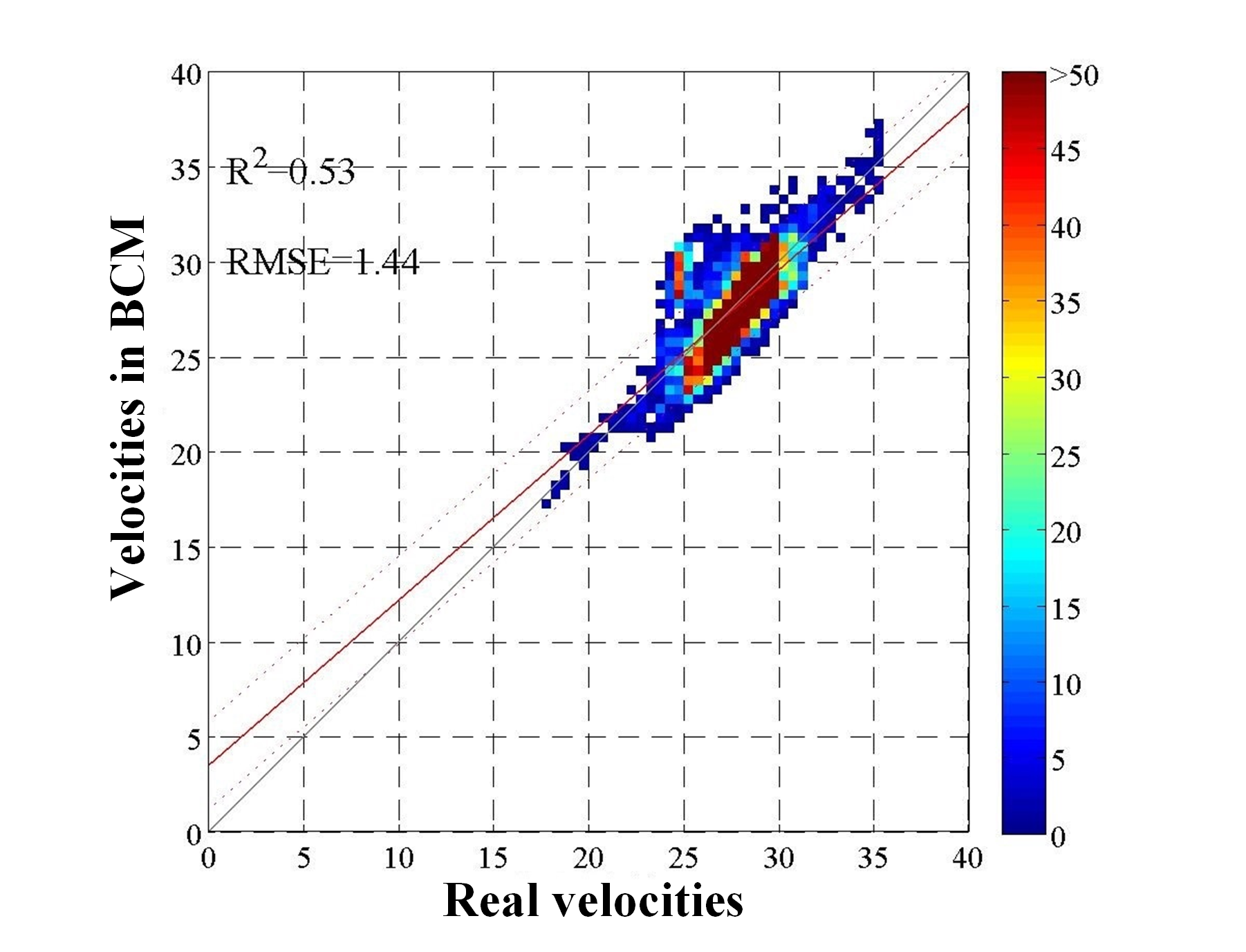}
		\label{k_1}
	}
	\subfigure[Real velocity and velocity in PBCM]{
		\includegraphics[scale=0.32]{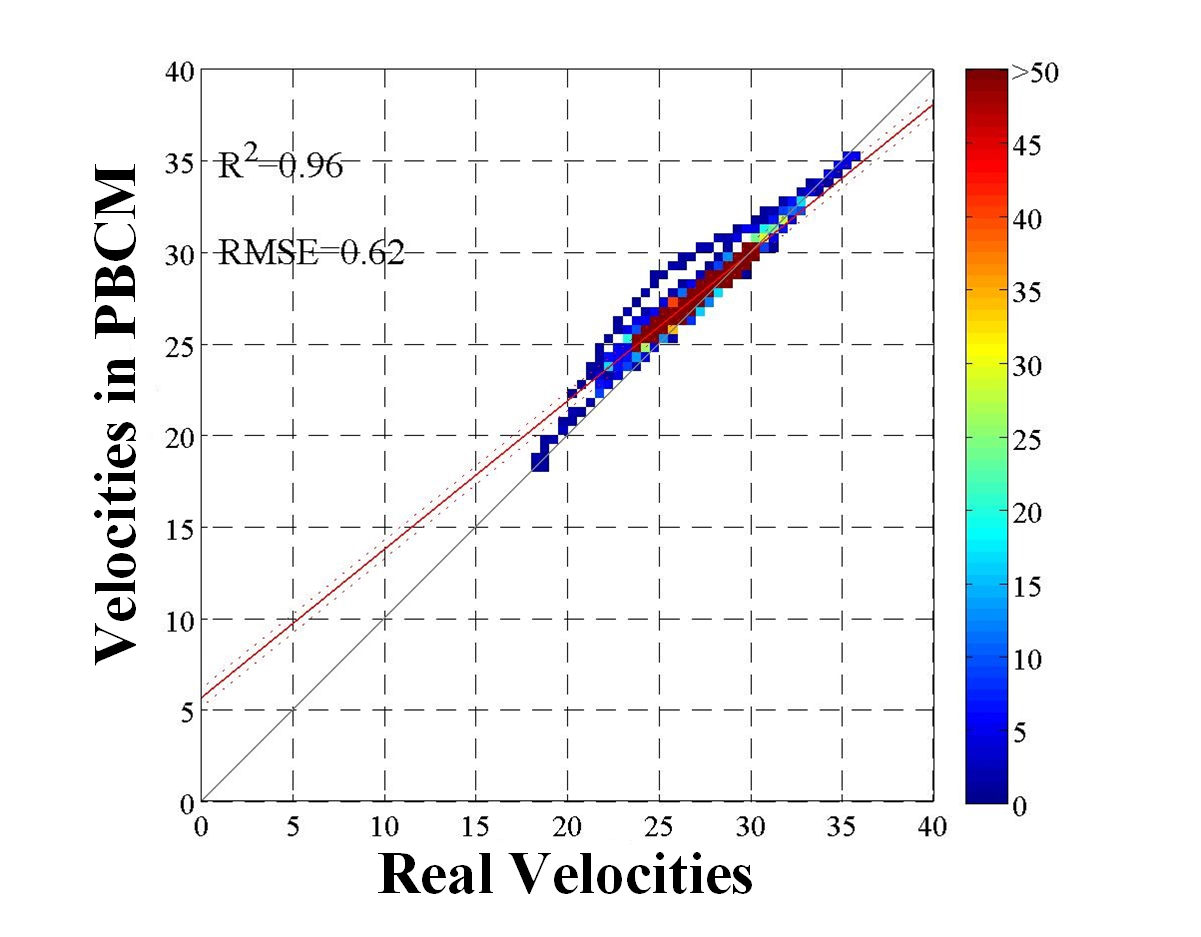}
		\label{k_2}
	}
	\subfigure[Real position and position in BCM]{
		\includegraphics[scale=0.28]{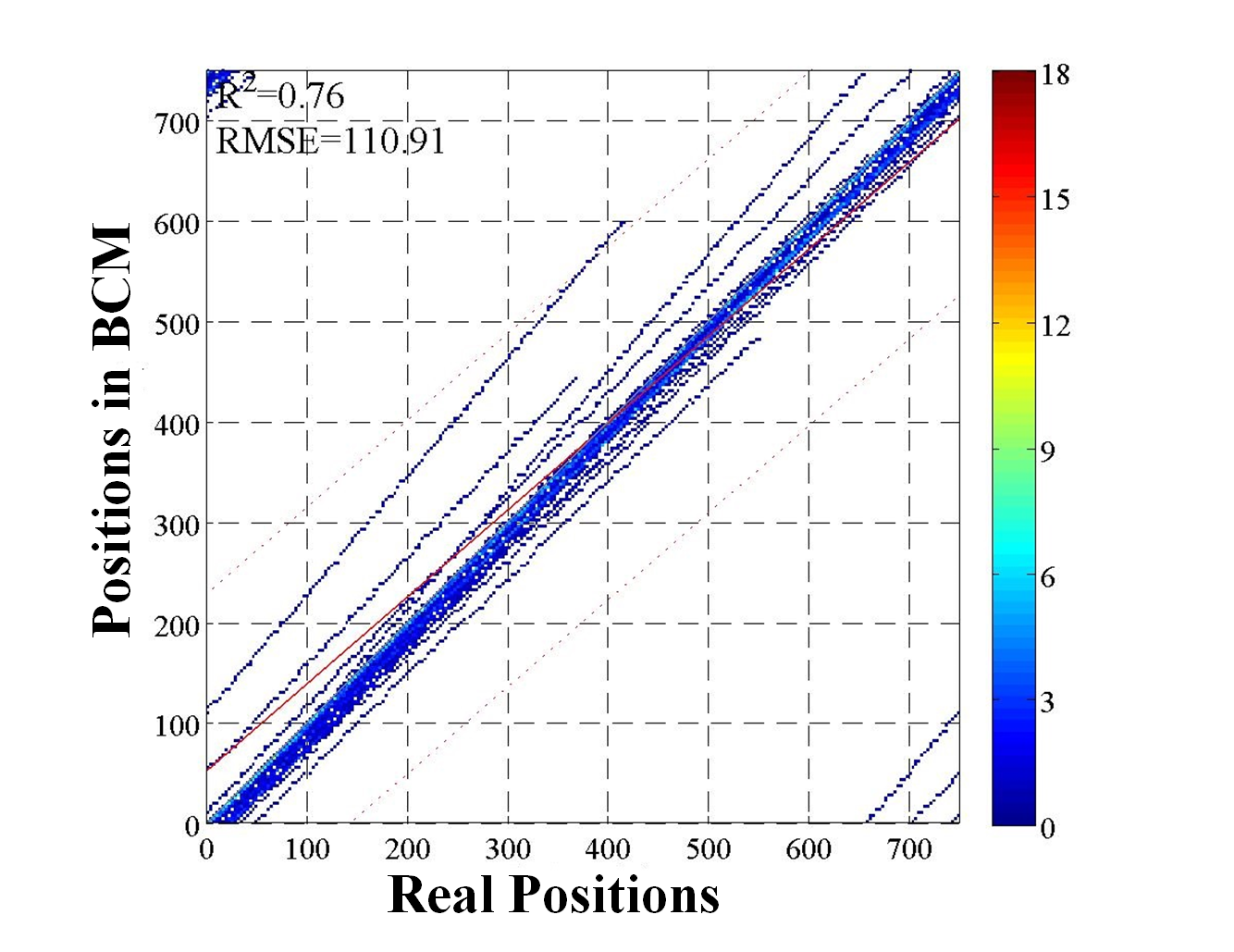}
		\label{k_3}
	}
	\subfigure[Real position and position in PBCM]{
		\includegraphics[scale=0.32]{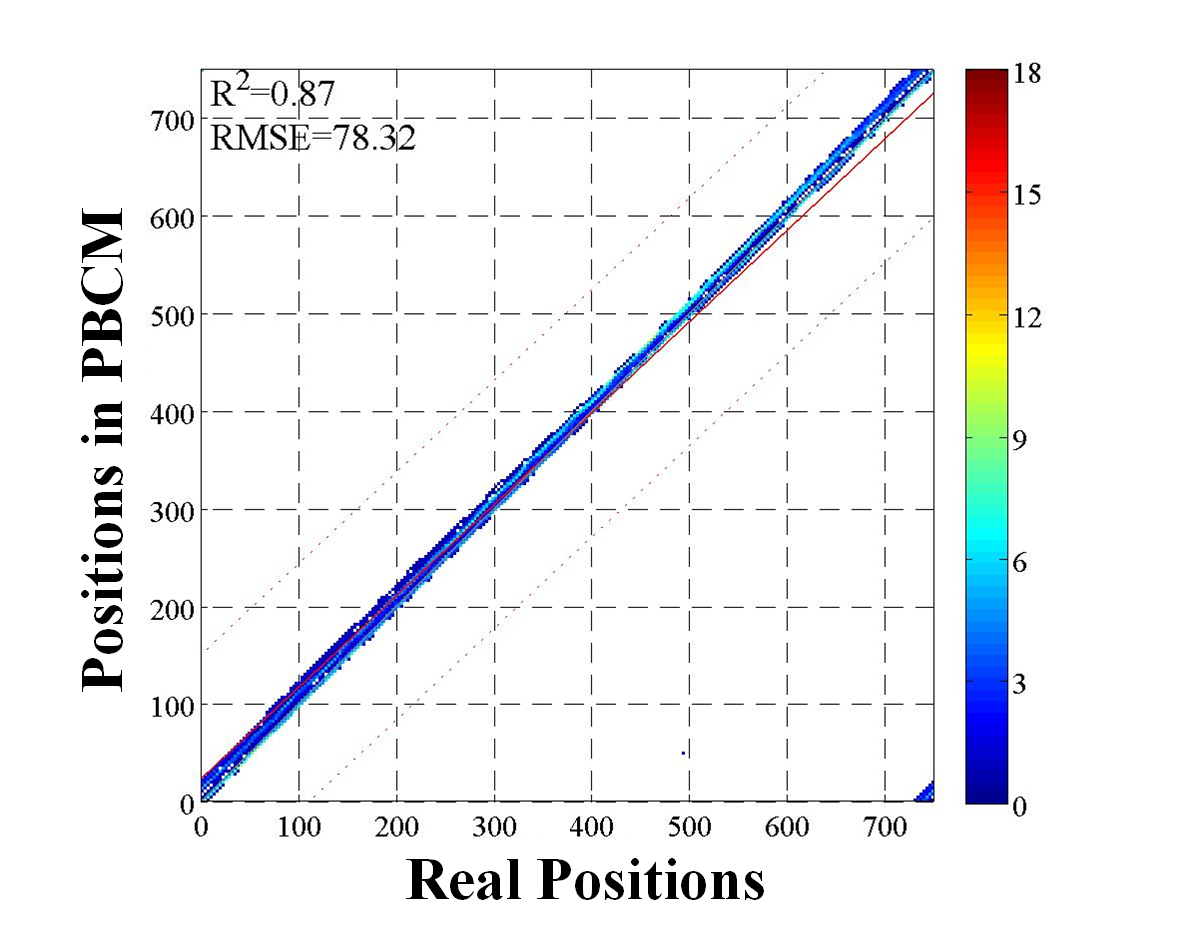}
		\label{k_4}
	}
	\caption{Error reduction}
	\label{pic_kal}
\end{figure}

As mentioned in section I, the uncertainties of vehicle measurements from on-board sensors are inevitable, and PBCM is introduced to alleviate the influence of inaccurate sensor information.We verified the effectiveness of the PBCM in eliminating errors by compare the $true$ values with the $measure$ values and $estimate$ values. In the real scenario, the $true$ values are not available, but in our simulation system, we can get the $true$ values at each time interval from simulation platform. By comparing the $measure$, $estimate$ and $real$ values, the performance of PBCM could be acquired, the results are shown in Fig.\ref{pic_kal}.

In case 15, 10\% measurement error was added to get the measurements. Position and velocity information is collected at each time intervals $t$ in the vehicle system, including the $measured$ values, $true$ values, and the $estimated$ values. The correlation coefficients $R^{2}$ and the root mean square error (RMSE) are calculated.

The Fig.\ref{k_1} and Fig.\ref{k_3} shows the difference between the $measured$ values and the $true$ values of vehicle velocity, Fig.\ref{k_2} and Fig.\ref{k_4} shows the difference between the $true$ values and the $estiamted$ values of vehicle position. Take the velocity for example, the $R^{2}$ rise from 0.53 to 0.96 which mean that the values after the estimation is more close to the real value, the $RMSE$ reduce from 1.44 to 0.62. The velocity results of other cases are listed in Tab.\ref{tab6}.
\begin{table}[H]
	\caption{Error reduction of PBCM}
	\label{tab6}
	\centering

	\begin{tabular}{@{}lllll@{}}
		\toprule
		& Case  & $R^{2}$ & $RMSE$ &  \\ \midrule
		& Case.13 & 0.75(0.98)  &  1.25(0.47)   &  \\
		& Case.14 & 0.73(0.97)  &  1.18(0.44)   &  \\
		& Case.15 & 0.53(0.96)  &  1.44(0.62)    &  \\
		& Case.16 & 0.51(0.94)  &  1.52(0.65)    &  \\ \bottomrule
	\end{tabular}
\end{table}

From the results above, it can be seen that by performing a estimation of the vehicle state, we could reduce the errors effectively, and the values we acquired is closer to the $real$ values in real scenario. This estimation gave us more accurate state of the system, then the estimate state could be applied to the predictable bilateral control systems effectively. 

\subsection{Models comparisons}
We already know that a vehicle system is considered more stable if vehicles in the system spaced on the road more evenly and the difference of vehicle velocity are more close\cite{trafficflowprove}. In this section, we will compare the performances of different models in two mainly aspect: velocity and position.

\subsubsection{Velocities in different model}
\begin{figure}
	\centering
	\subfigure[The velocity of each vehicle at 9s of different model]{
		\includegraphics[scale=0.16]{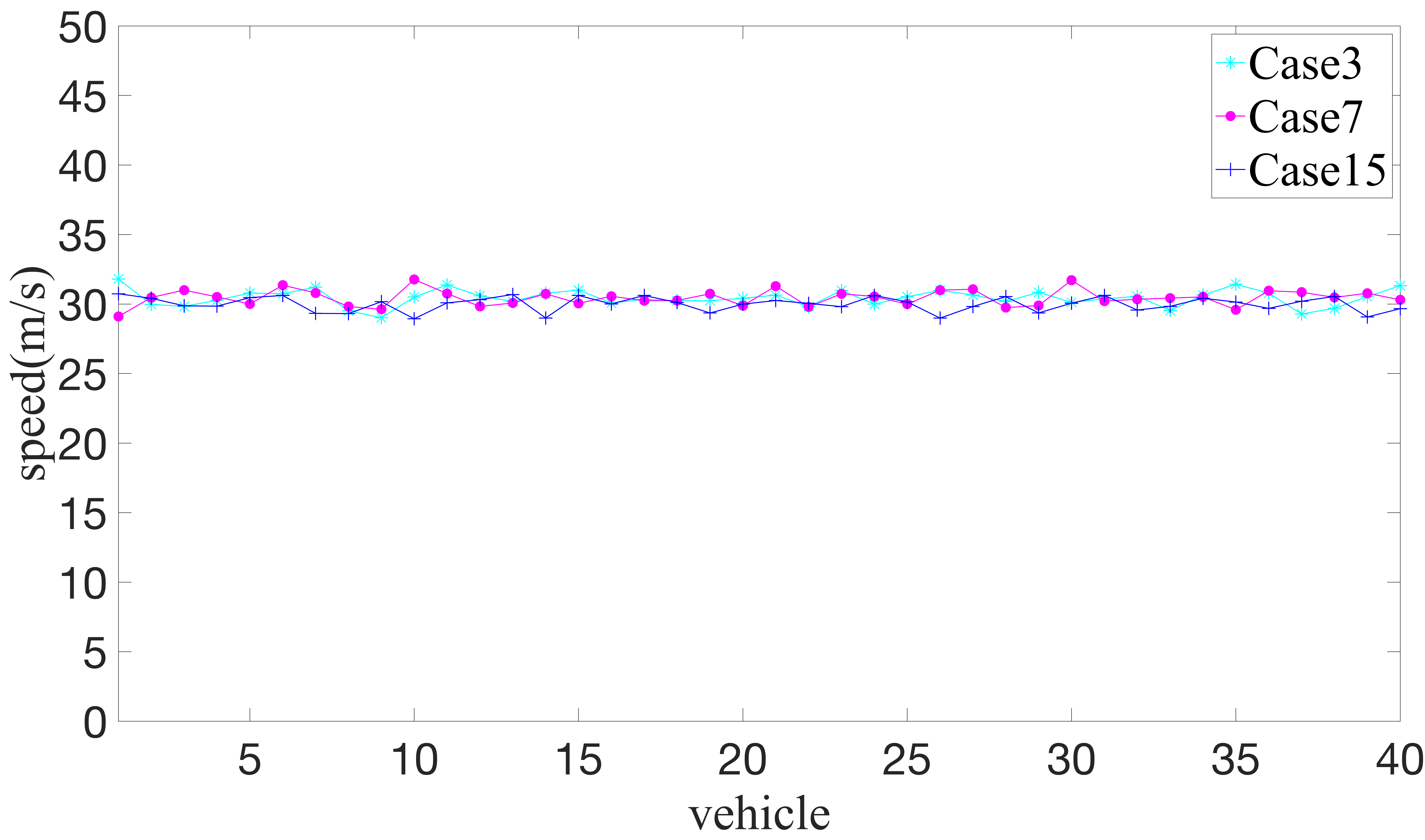}
		\label{cf9}
	}
	\subfigure[The velocity of each vehicle at 30s of different model]{
		\includegraphics[scale=0.16]{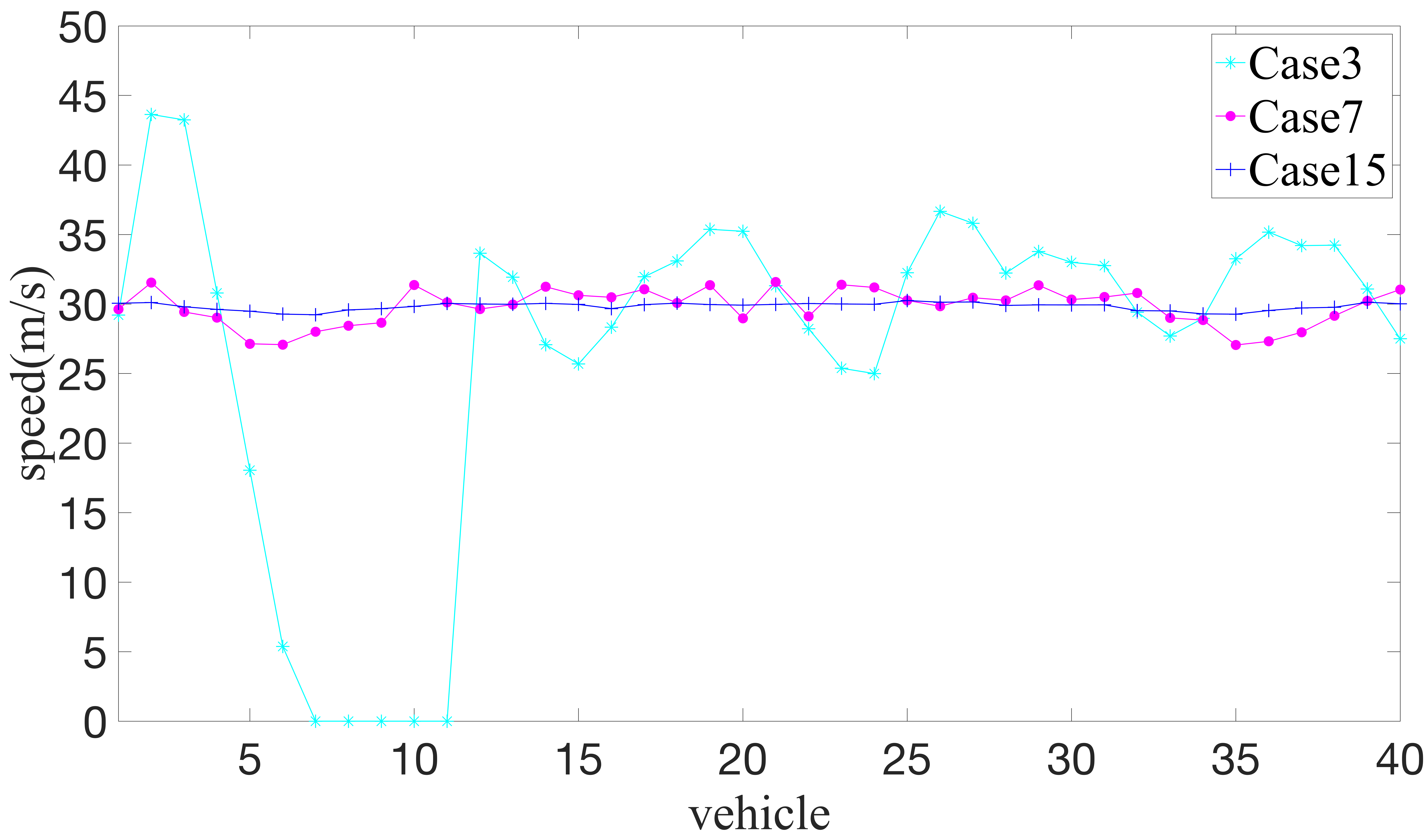}
		\label{cf30}
	}
	\subfigure[The velocity of each vehicle at 100s of different model]{
		\includegraphics[scale=0.16]{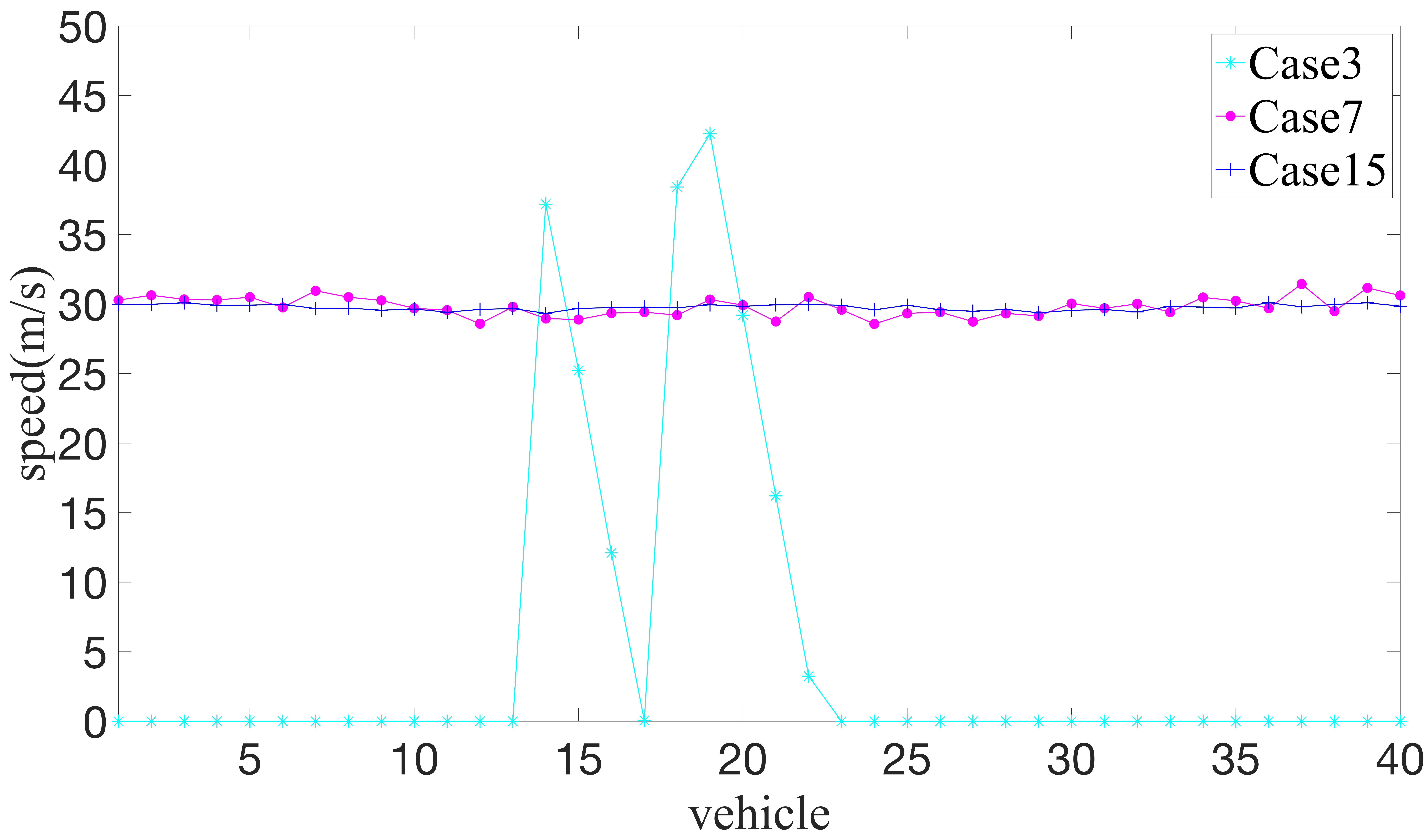}
		\label{cf100}
	}
	\subfigure[The velocity of each vehicle at 300s of different model]{
		\includegraphics[scale=0.16]{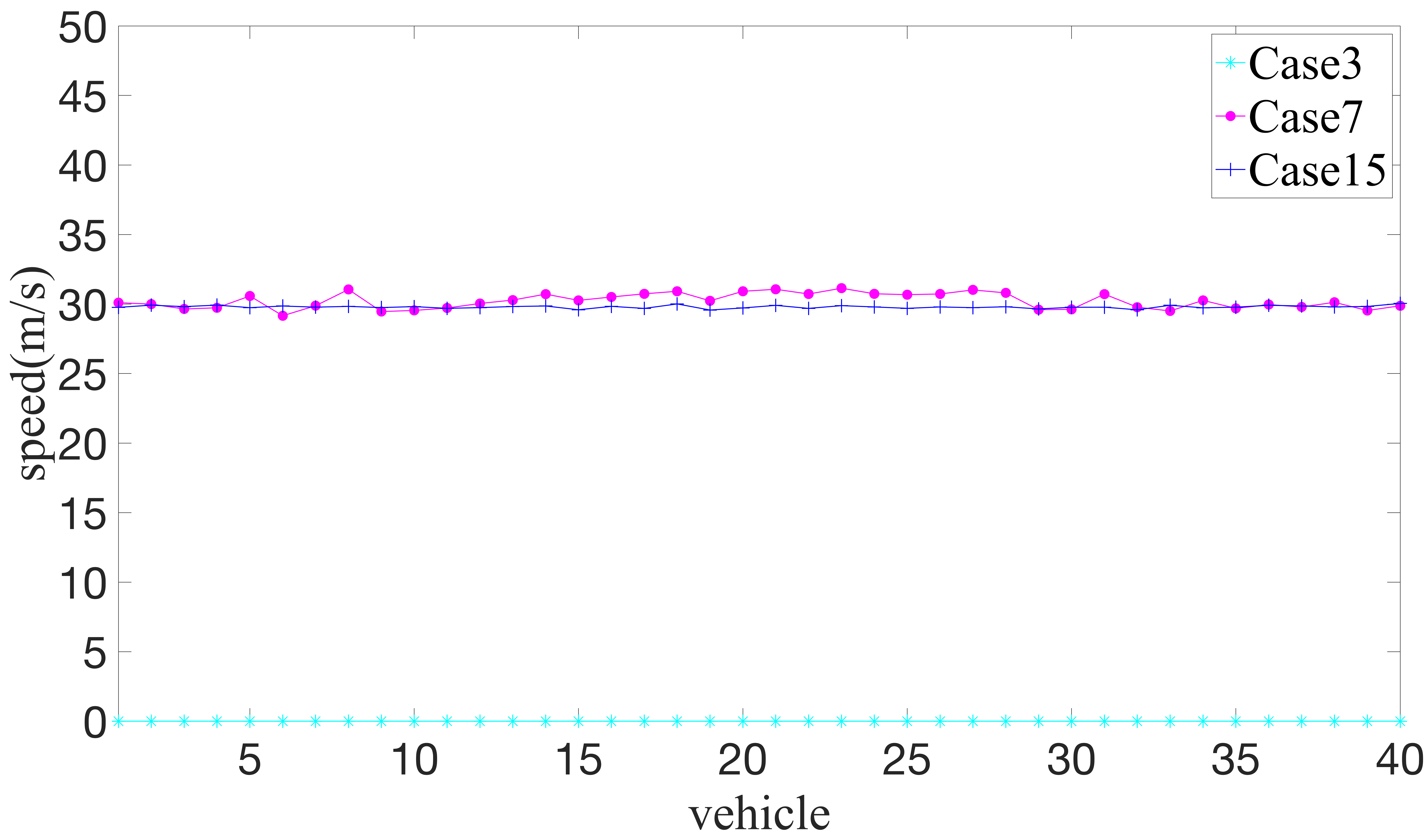}
		\label{cf300}
	}
	\caption{Change of the vehicle velocity in the traffic flow with time}
	\label{pic_vs}
\end{figure}

Fig.\ref{pic_vs} is the snapshots of each vehicle's speed at different time under three different models. Here, in these scenarios, case3, case7 and case15 are extracted for example, the delay is set to 0.1$s$ and the measurement error is set to $10\%$. During the first 10$s$, the vehicles are  evenly spaced and moving approximately at the same speed as shown in Fig.\ref{cf9}, then, the vehicle of index 5 braked barking at 10$s$ last for 2$s$, after that, typical "stop-and-go" traffic pattern appears pretty soon within FM. Fig.\ref{cf30} shows the speed of each vehicle under different control model, and it can be seen from the figures that the vehicles under FM has cause traffic congestion(i.e the velocities of vehicle 6 to vehicle 10 are approximately close to 0), and FM still not working at 100$s$ and almost all vehicles stopped, after 300 seconds, the system is totally not working under FM and all vehicles occurred severe congestion (see in Fig.\ref{cf30}\ref{cf100}). And it also can be seen that the BCM and the PBCM have a better performance, and the PBCM is slightly better than the BCM (the RMSE of BCM and PBCM are 0.23 and 0.17 at 100$s$) in terms of velocity.
 
From Fig.\ref{cf30}-Fig.\ref{cf300} the FM gradually became not working, and show the worst performance compare with the BCM and PBCM, therefore in the remaining experiments, we would focus on the comparison of BCM, MBCM and PBCM. 

\begin{figure}
	\centering
	\subfigure[The velocities in PBCM]{
		\includegraphics[scale=0.15]{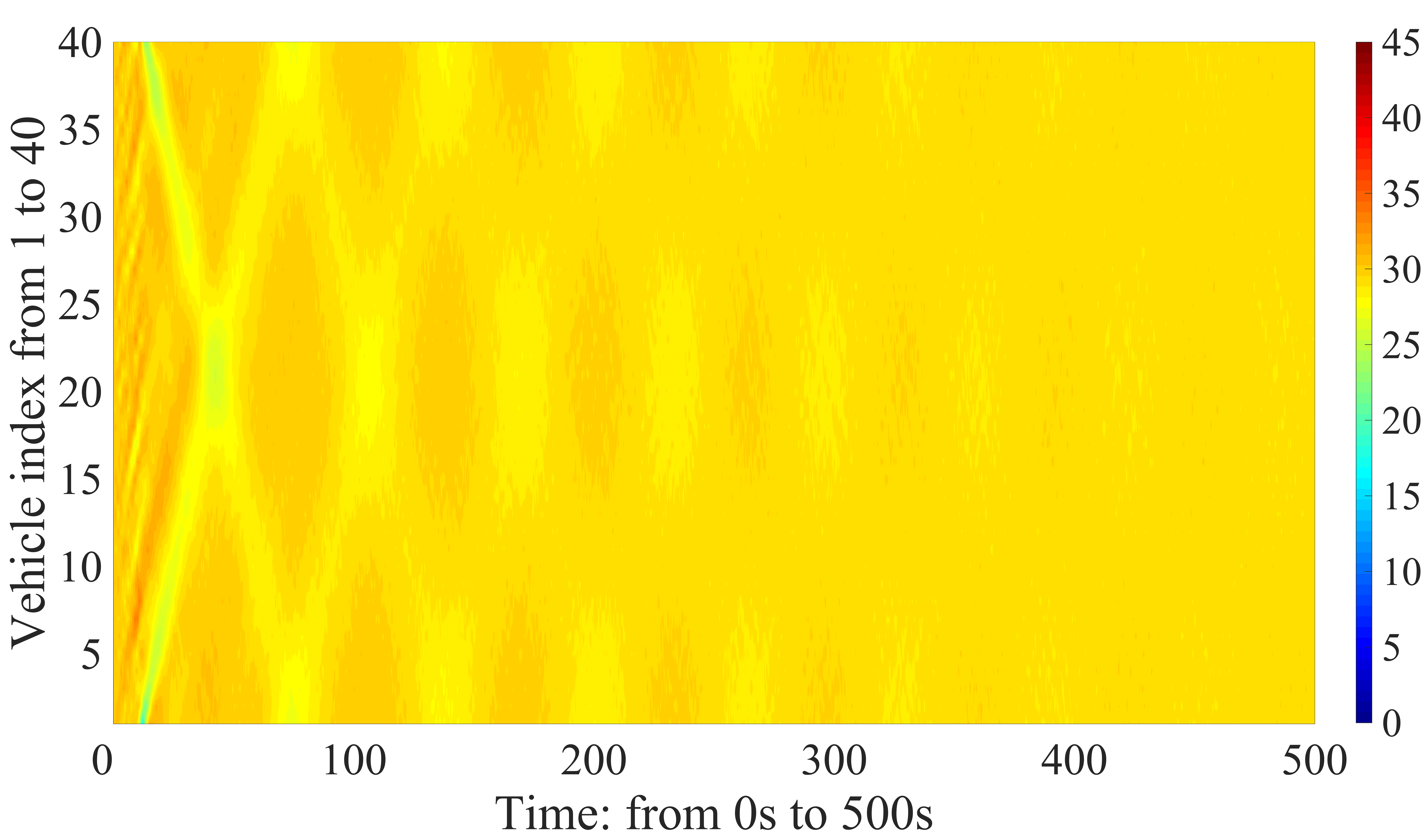}
		\label{novel}
	}
	\subfigure[The velocities in MBCM]{
		\includegraphics[scale=0.15]{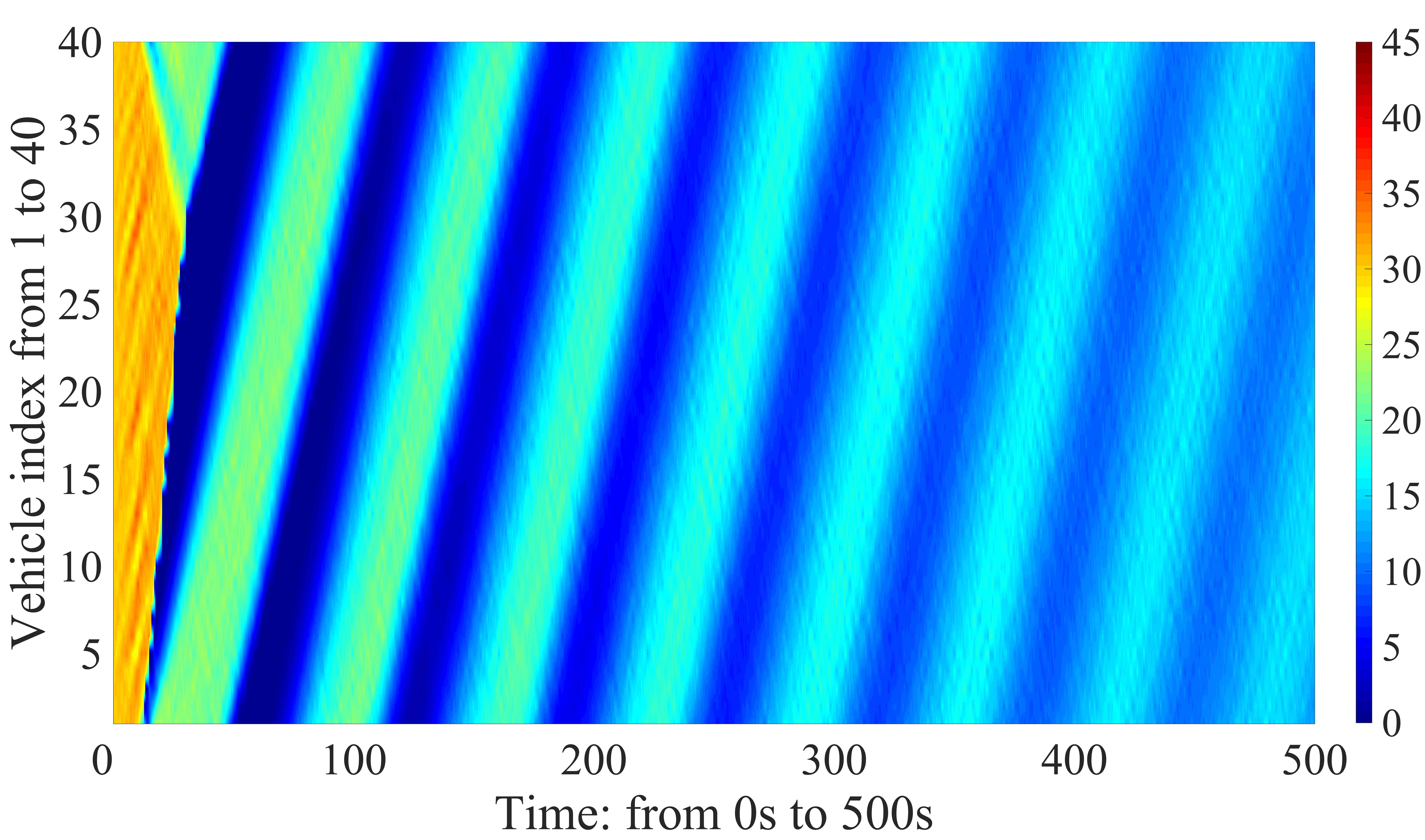}
		\label{tr_bi}
	}
	\subfigure[The velocities in BCM]{
		\includegraphics[scale=0.15]{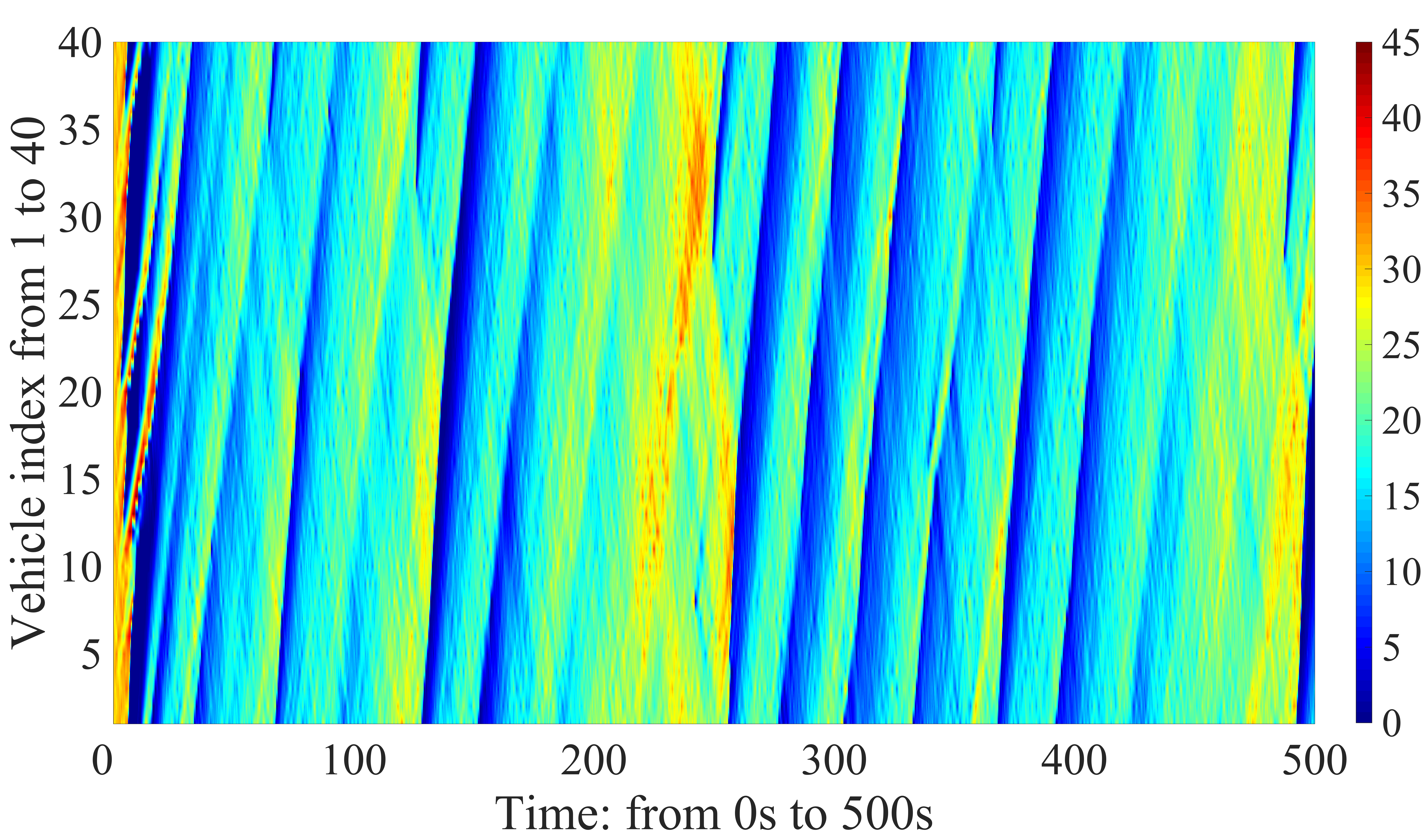}
		\label{multi_bi}
	}
	\caption{Velocity heat Map}
	\label{heatmap}
\end{figure}

In addition, vehicles speed in the whole system is depicted in the form of a heat map. Heat map can well indicate the stability of system parameters during the entire operation of the system. The results are shown in Fig.\ref{heatmap}. The x-axis represents time, and the y-axis represents the index of vehicle. Different colors represent different vehicle speed, the vehicle system could be considered more stable if the tone of different vehicle is close which means the vehicle difference among vehicles is small and the more bright color represent higher velocity which indicate our system run more efficient.

All cases begin with the FM and velocity of each vehicle is 30$m/s$, so the tone in the heat map is yellow at the beginning. At the first 10s, all vehicles work under the FM and velocities are reducing gradually, when the braking happened, as the time going, the velocities in BCM as shown in Fig.\ref{multi_bi} is reducing gradually and stabilized around $25m/s$. In MBCM, because the current vehicle need more information rather just the PV and the SV, the V2V communication time delay will be bigger so the speed of vehicles within MBCM is more lower (about 18$m/s$), however the PBCM always have higher velocities (around 30$m/s$) and small velocity difference as shown in \ref{novel}.

From the comparison of velocities, it could be seen that whether it is directly sampling and drawing the speed during operation, or expressing the overall speed change of the entire system in the form of a heat map, the PBCM has better performance than other models.

\subsubsection{Performances on vehicle's distance}
In this subsection, we choose three different measurement parameters to measure the distance between vehicles in the vehicle system. We use the $MinDinstance$, $MaxDistance$ and the error bar, we define:
$$
\begin{array}{l}
	\text { MinDistance }=\left(d \min _{1}+d \min _{2}+d \min _{3}\right) / 3 \\
	\text { MaxDistance }=\left(d \max _{1}+d \max _{2}+d \max _{3}\right) / 3
\end{array}
$$
where $d \min _{i}$ and $d \max_{i}$ represent the $i$th minimum and maximum value among the all vehicle distances for each time t.

\begin{figure}
	\centering
	\subfigure[]{
		\includegraphics[scale=0.14]{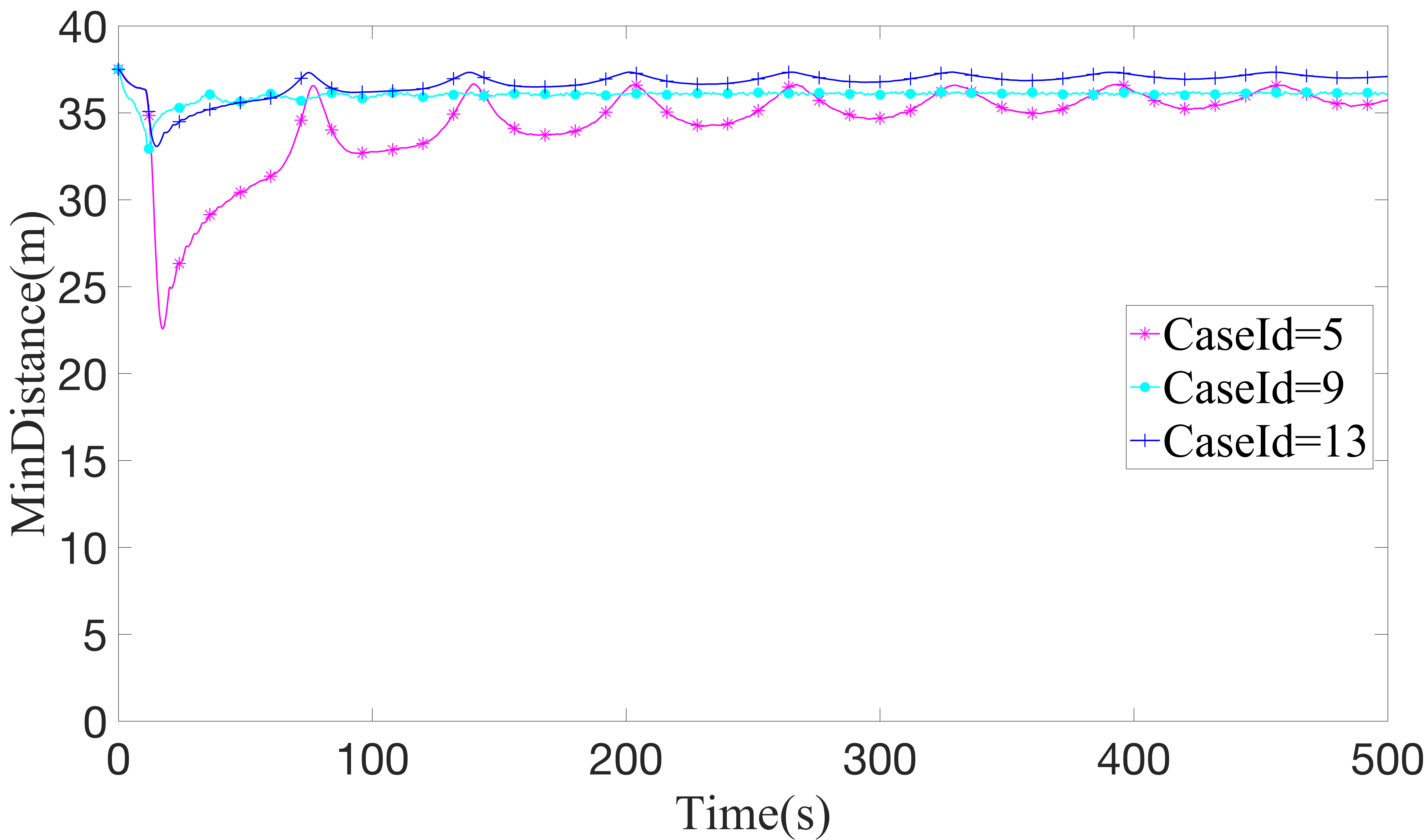}
		\label{bc_a}
	}
	\subfigure[]{
		\includegraphics[scale=0.14]{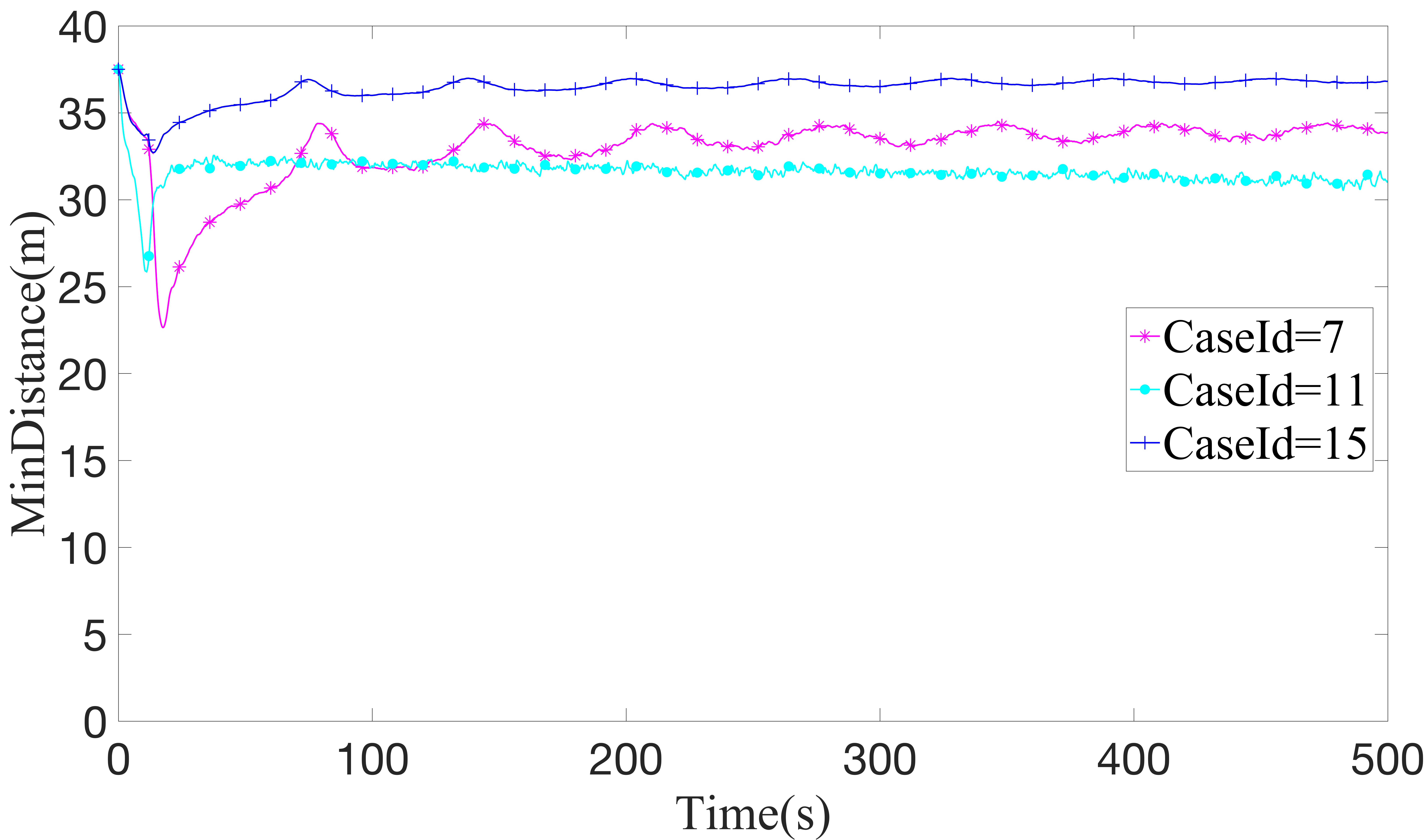}
		\label{bc_b}
	}
	\subfigure[]{
		\includegraphics[scale=0.14]{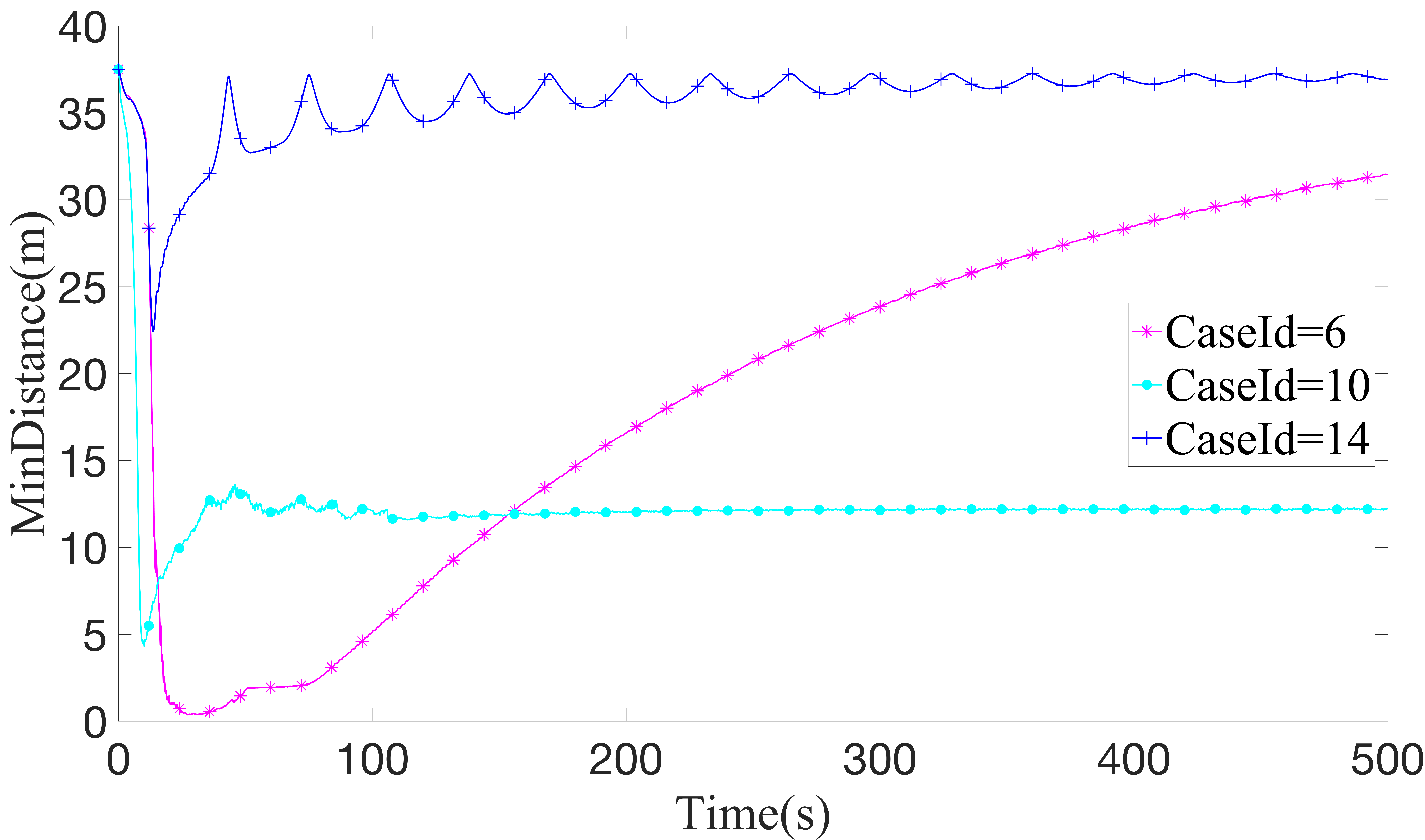}
		\label{bc_c}
	}
	\subfigure[]{
		\includegraphics[scale=0.147]{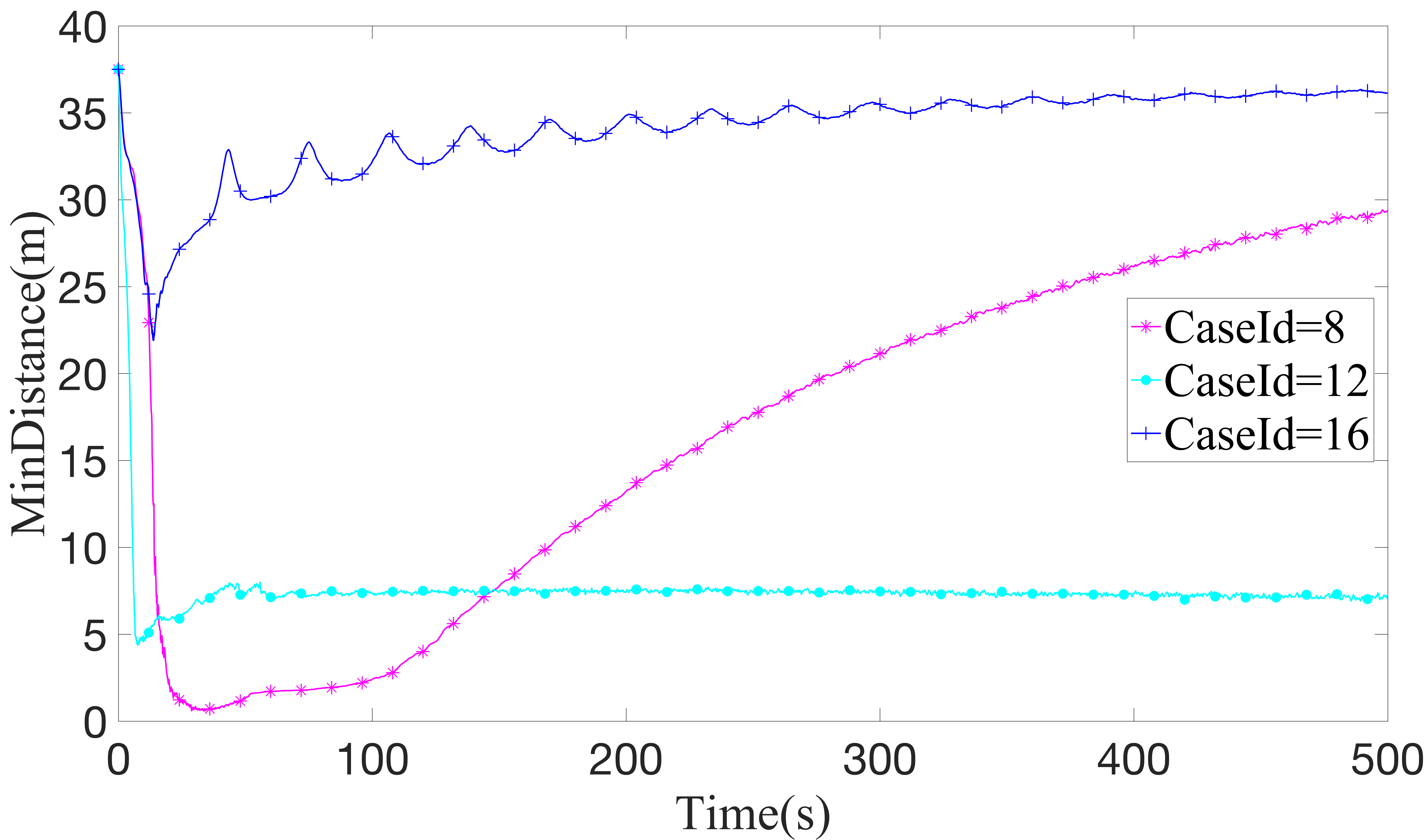}
		\label{bc_d}
	}
	\caption{Change of the minimum vehicle distance in the traffic flow with time}
	\label{bc_group}
\end{figure}

The $MinDinstance$ of different models with different errors and delay is shown in Fig.\ref{bc_group}. The total length of the road is 1500$m$, Thus if the all vehicles distributed on the road evenly, the $MinDistance$ and $MaxDistance$ will be the same (i.e 37.5$m$) but because of the errors and delay, vehicles will never distributed on the road evenly, the $MinDistance$ will smaller than 37.5$m$ and the $MaxDistance$ will be bigger than 37.5$m$, the bigger/smaller the $MinDistance$/$MaxDistance$ is the better the model performed.

Firstly, for each figure in Fig.\ref{bc_group}, the PBCM always have the best performance than other model. Take the Fig.\ref{bc_b} for example, when the braking happen at 10$s$, the $MinDistance$ within all models reduced rapidly (i.e PBCM reduce from 35m to 33m, MBCM reduce from 35 to 26$m$, BCM reduce from 35m to 24$m$), but the PBCM shows the smallest reduction which means the PBCM can resist the disturbance than BCM and MBCM.

Secondly, we explored the impact of errors on different models, as shown in Fig.\ref{bc_a} and Fig.\ref{bc_b}, in these two scenarios, the error is set to 3\% and 10\%, and the delay is set to 0.1$s$. We could find that when the error increases from 3\% to 10\%, the change of the blue curve (PBCM) is not obvious and the $MinDistance$ was settled at about 36$m$ finally, while the red curve (BCM) and the cyan curve (MBCM)  both decrease rapidly and the $MinDistances$ are settled at about 31$m$ and 33$m$
finally. In Fig.\ref{bc_c} and Fig.\ref{bc_d}, the error is 3\% and 10\%, respectively, and the delay is 0.2s. The change of each curve after the error increases is the same as that of Fig.\ref{bc_a} and Fig.\ref{bc_b}. This set of comparative experiments produced similar results. This proved that compared with the other two models, PBCM can eliminate the influence of error better.

Thirdly, we compared the performance of different model under different delay, we consider two sets of experiments. In the first group, we set the error to 3\% and the delay to 0.1s and 0.2s respectively. The results are shown in Fig.\ref{bc_a} and Fig.\ref{bc_c}. In the second group, we set the error to 10\% and the delay to 0.1s and 0.2s respectively. The results are shown in Fig.\ref{bc_b} and Fig.\ref{bc_d}. First, both sets of experimental results show that when the delay increases, the $MinDinstance$ within BCM and the MBCM decreases sharply but the $MinDistance$ within PBCM just reduce from 36$m$ to 35$m$. Second, the $MinDistance$ of the BCM toke about 500$s$ to stabilize, although the MBCM can quickly restore to the stable state, but the $MinDinstance$ was reduced by about 60\%-75\% compared to the initial state, in contrast, the PBCM could recover to the initial state of more quickly (about 50$s$).

\begin{figure}
	\centering
	\subfigure[]{
		\includegraphics[scale=0.17]{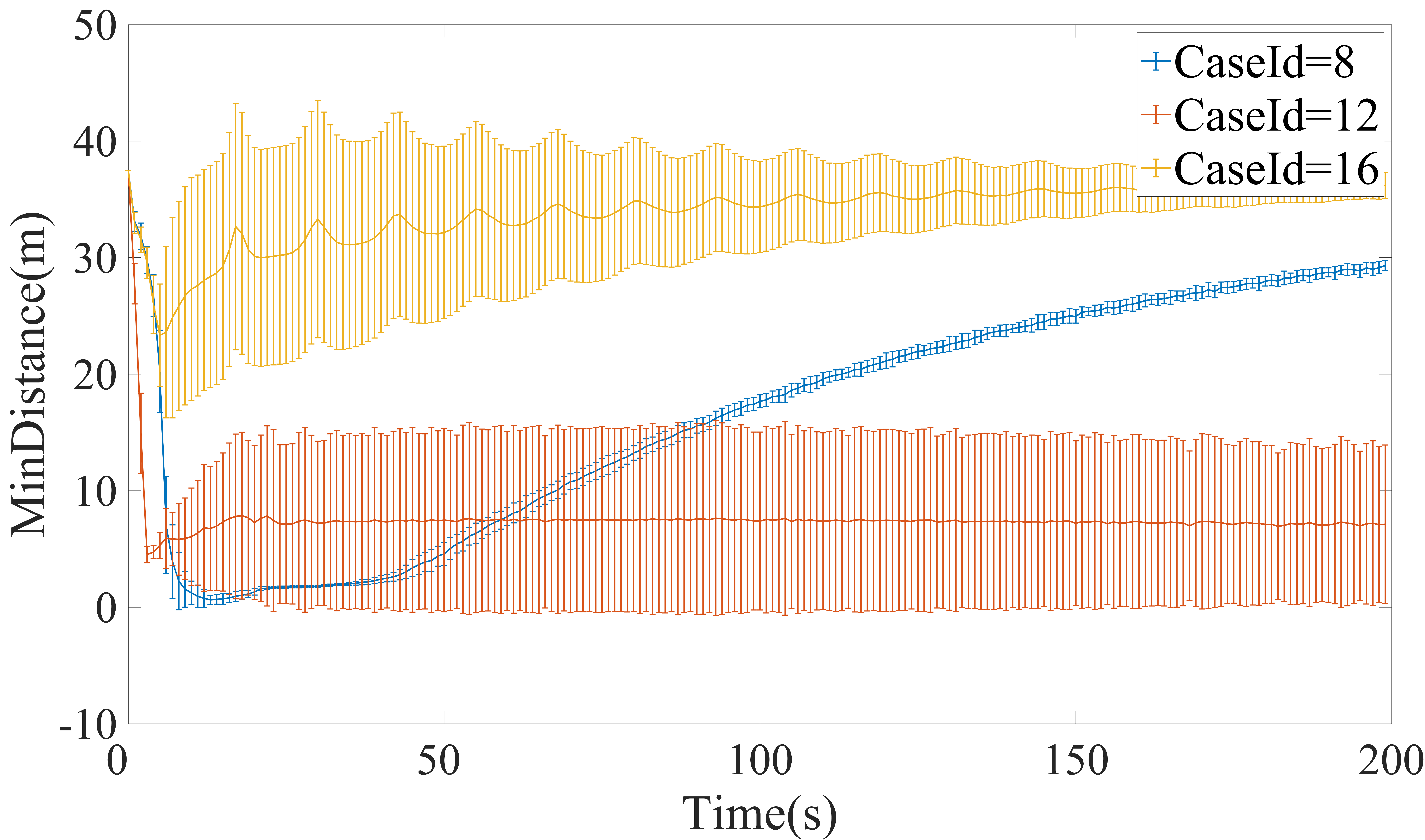}
		\label{d_a}
	}
	\subfigure[]{
		\includegraphics[scale=0.17]{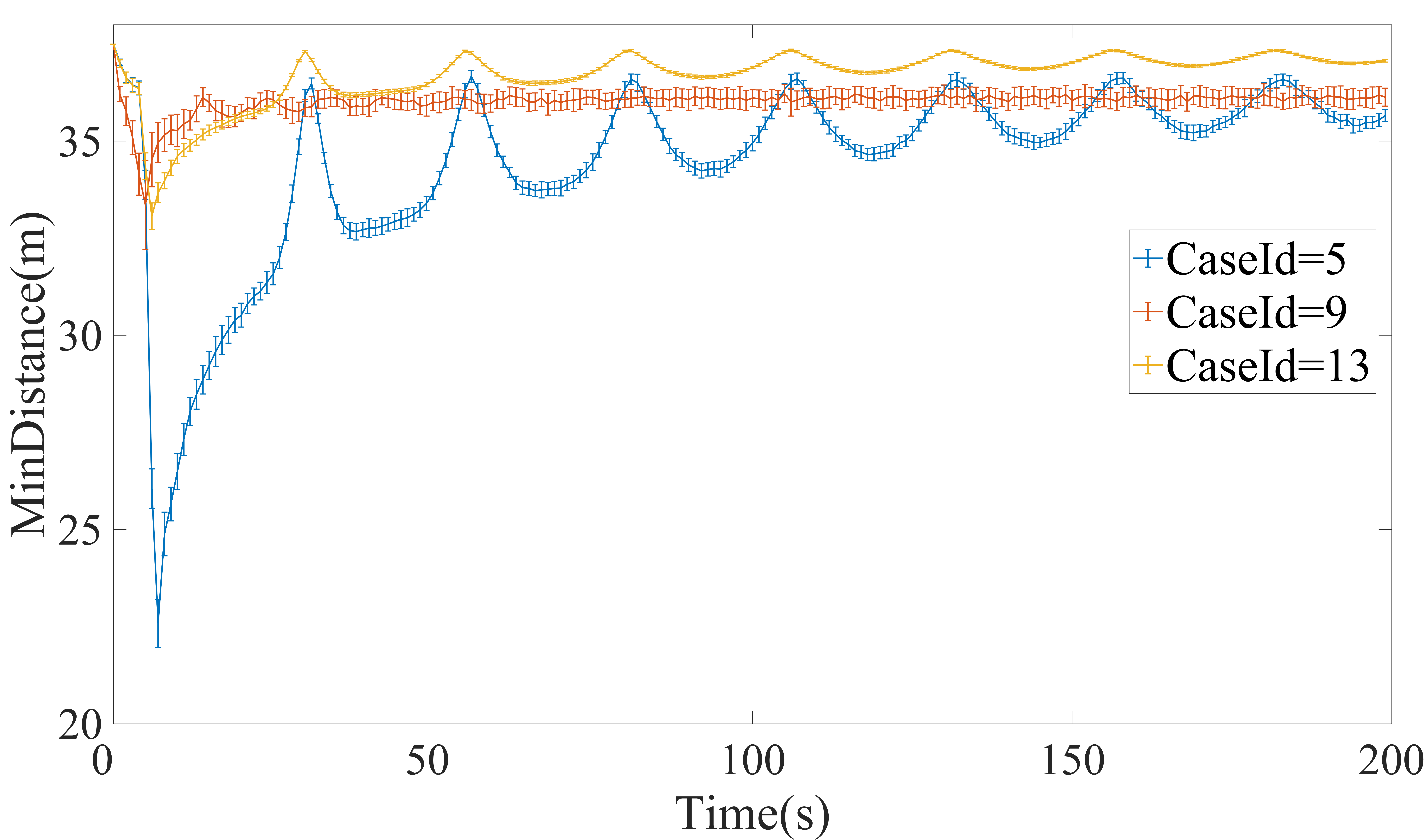}
		\label{d_b}
	}
	\caption{The error bar of minimum distance}
	\label{d_group}
\end{figure}

In addition, the error bar of the minimum distance is shown Fig.\ref{d_group}, the error bar can represents the uncertainly (RMSE) of related data.
it showed the performances of PBCM and BCM and MBCM under different delay and errors. Compare with Fig.\ref{d_a} and Fig.\ref{d_b}, it can be easily get that whether it is larger error, larger delay (see Fig.\ref{d_a}), or smaller error and smaller delay(as shown in Fig.\ref{d_b}), the PBCM always has small error bar and show better performance than the other two models.

\begin{figure}
	\centering
	\subfigure[]{
		\includegraphics[scale=0.17]{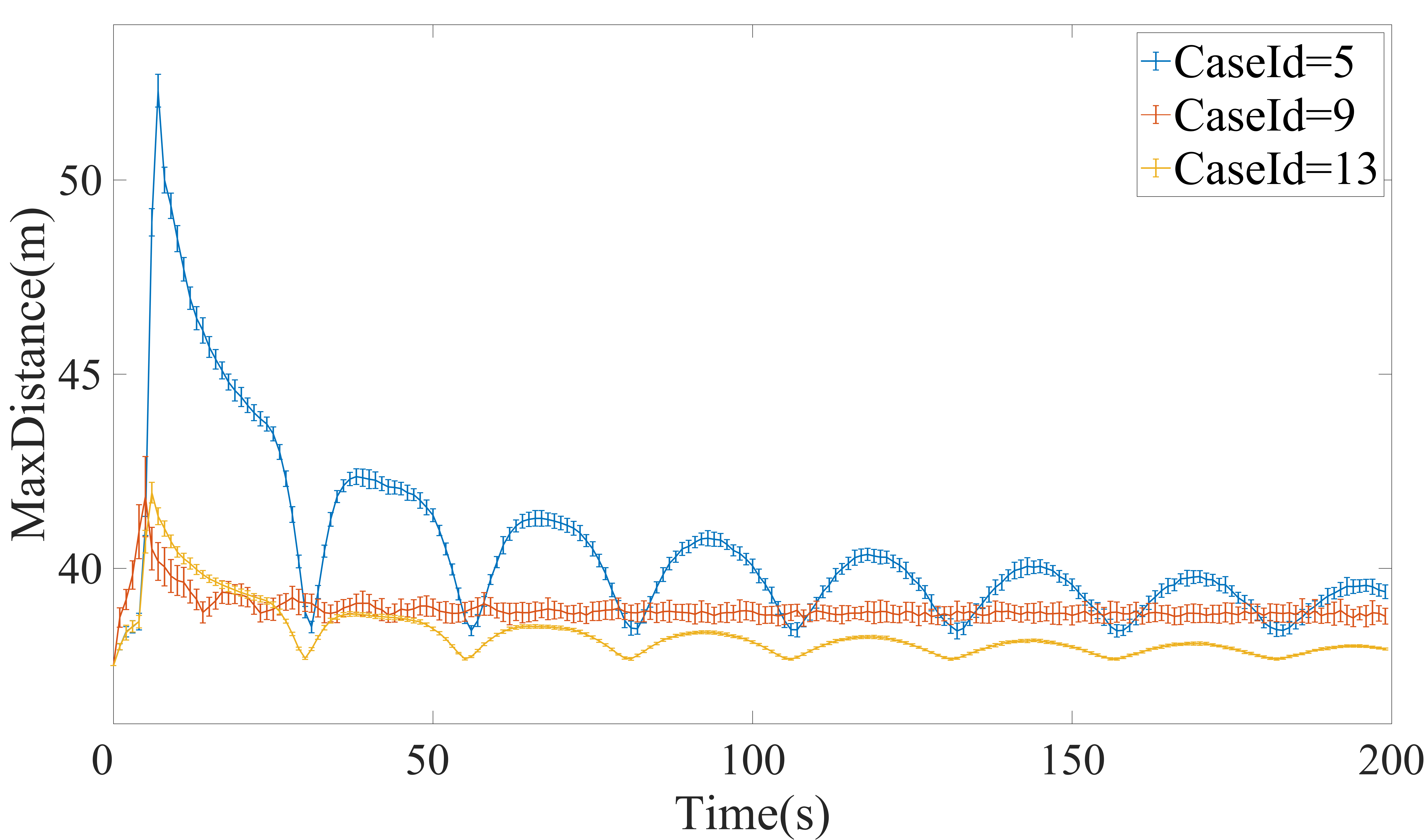}
		\label{bc_a_a}
	}
	\subfigure[]{
		\includegraphics[scale=0.17]{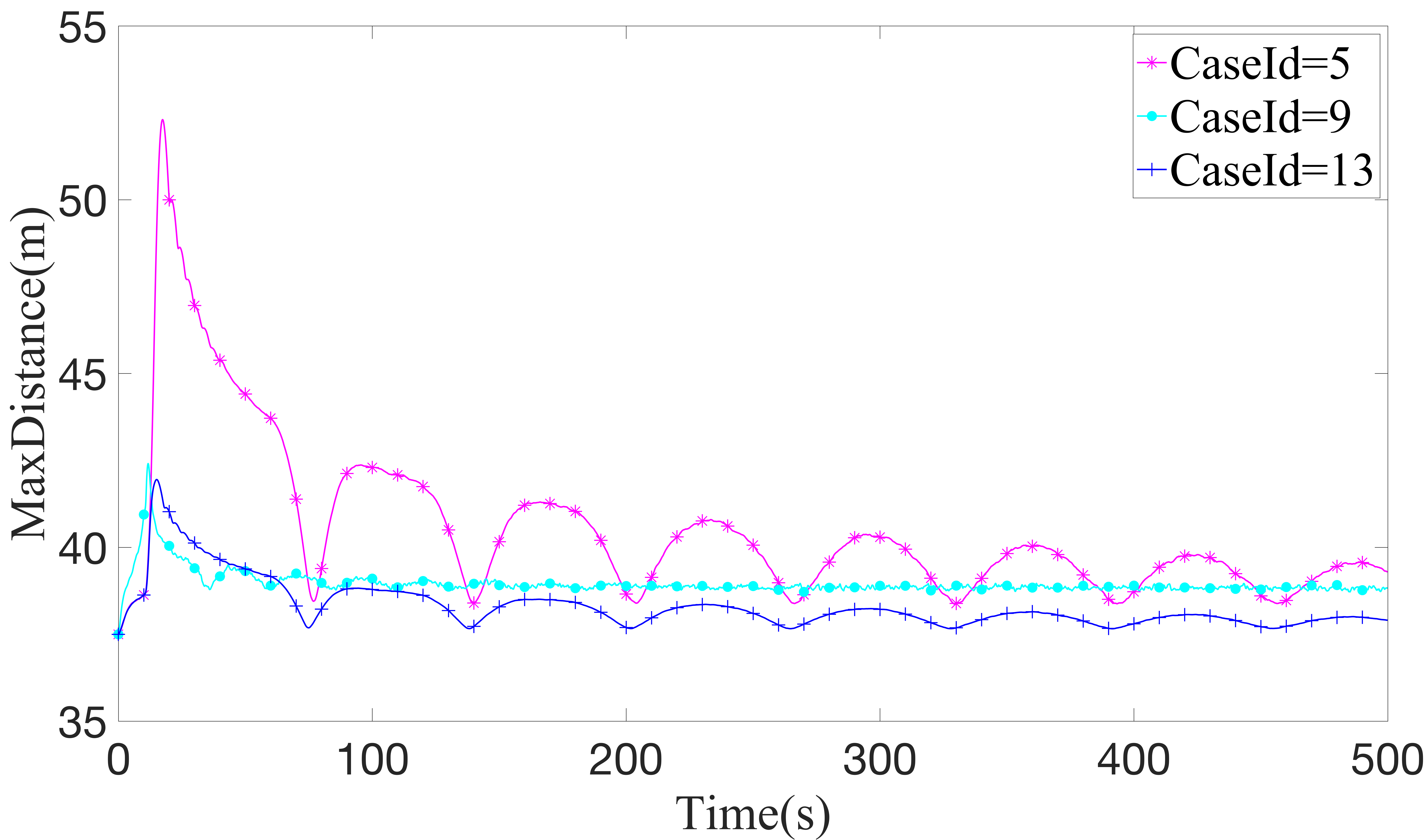}
		\label{bc_b_b}
	}
	\caption{Max distance and error bar of different model}
	\label{bc_group_g}
\end{figure}

Similarly, the maximum distance (Fig.\ref{bc_a_a}) and the error bar of maximum distance (Fig.\ref{bc_b_b}) of vehicles during the all running process is shown in Fig.\ref{bc_group_g}.

Similar to the $MinDinstance$, the PBCM always have the smallest $MaxDistance$(around 38$m$) than MBCM and BCM, and the MBCM takes more vehicles into consideration, so  the MBCM convergence faster than BCM, but because more vehicles are considered and the information was transmit with V2V, so the delay also increased and the performance is not as well as PBCM.

From the result above, it could be concluded that with the Kalman-based model to reduce the errors and prediction procedure to eliminate the influence of delay PBCM have a better performance than the BCM and MBCM within real scenarios, it could reduce the instability effectively.

\section{Conclusions}
In this paper, a real-time and reliable model based on the BCM called PBCM is presented. The aim of our work mainly contain two parts: First, with the development of on-vehicle detection technology, the collection of vehicle information is more diversified, thus the uncertainties of vehicle information are explored. Kalman-base model is introduced to reduce the error of vehicle state. Second, each vehicle will communicate with the preceding vehicle and succeeding vehicle, so the communication delay and the system delay are both take into consideration, a state prediction procedure is proposed to alleviate the influence of delay. With the comparison of different models (i.e FM, BCM, MBCM, PBCM), it is shown that PBCM which consider both the delay and error always keeps high performance in the simulations.

Our model could be further improved in some aspects. First, our experiments are based on a simulation platform, further field experiments in real scenarios and develop common modules contain both radar, V2V module and the PBCM model should be carried out in the future. Second, in the state prediction part, Kalman-based model was adopted. In recent years, lots of prediction models have been proposed in machine learning, how to apply the machine learning methods into the state prediction part will be further explored.
\bibliographystyle{IEEEtran}
\bibliography{references/references}
\begin{IEEEbiography}[{\includegraphics[width=1in,height=1.25in,clip,keepaspectratio]{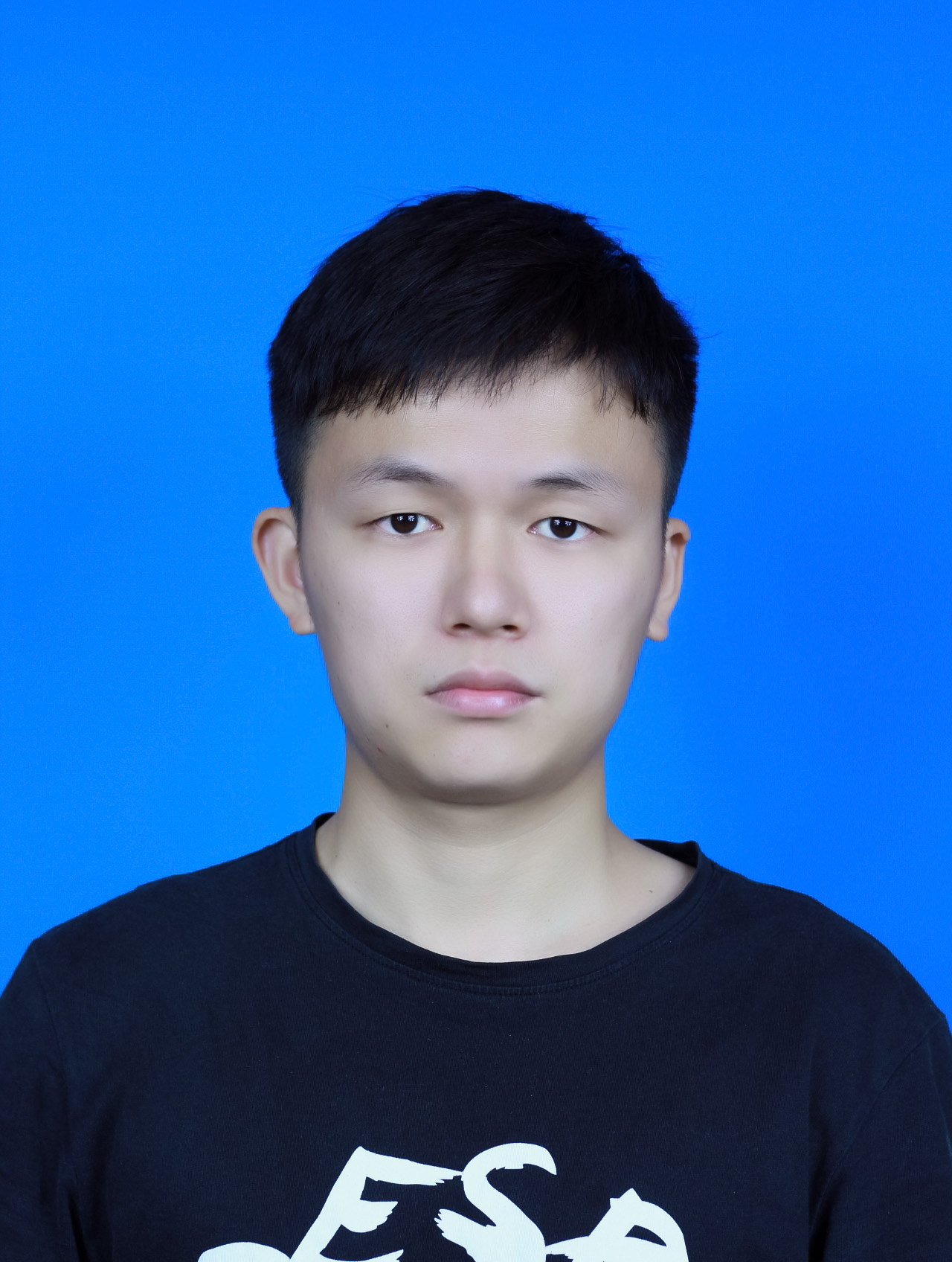}}]{Jiancheng Fang}
received the B.E. degree from the University of Electronic Science and Technology of China, Chengdu, China, in 2018, where he is currently pursuing the M.S. degree in computer science.

His current research interests include the Internet of Things and ITS.
\end{IEEEbiography}
\begin{IEEEbiography}[{\includegraphics[width=1in,height=1.25in,clip,keepaspectratio]{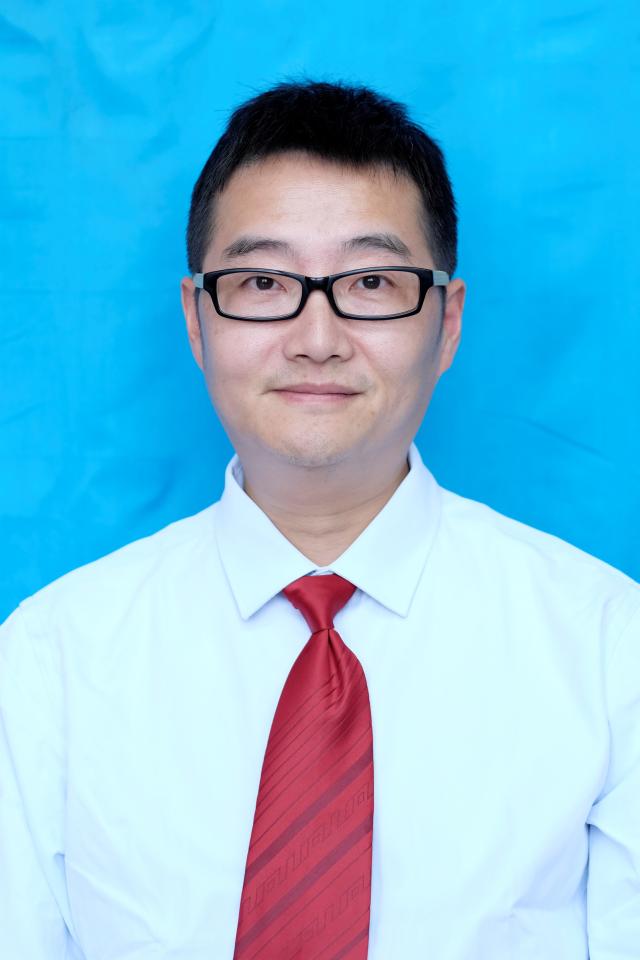}}]{Yu Xiang}
YU XIANG (Member, IEEE) received the B.S., M.S., and Ph.D. degrees from the University of Electronic Science and Technology of China (UESTC), Chengdu, Sichuan, China, in 1995, 1998, and 2003, respectively. He joined the UESTC, in 2003, and became an Associate Professor, in 2006. From 2014 to 2015, he was a Visiting Scholar with The University of Melbourne, Australia. 

His current research interests include computer networks, intelligent transportation systems and deep learning.
\end{IEEEbiography}
\begin{IEEEbiography}[{\includegraphics[width=1in,height=1.25in,clip,keepaspectratio]{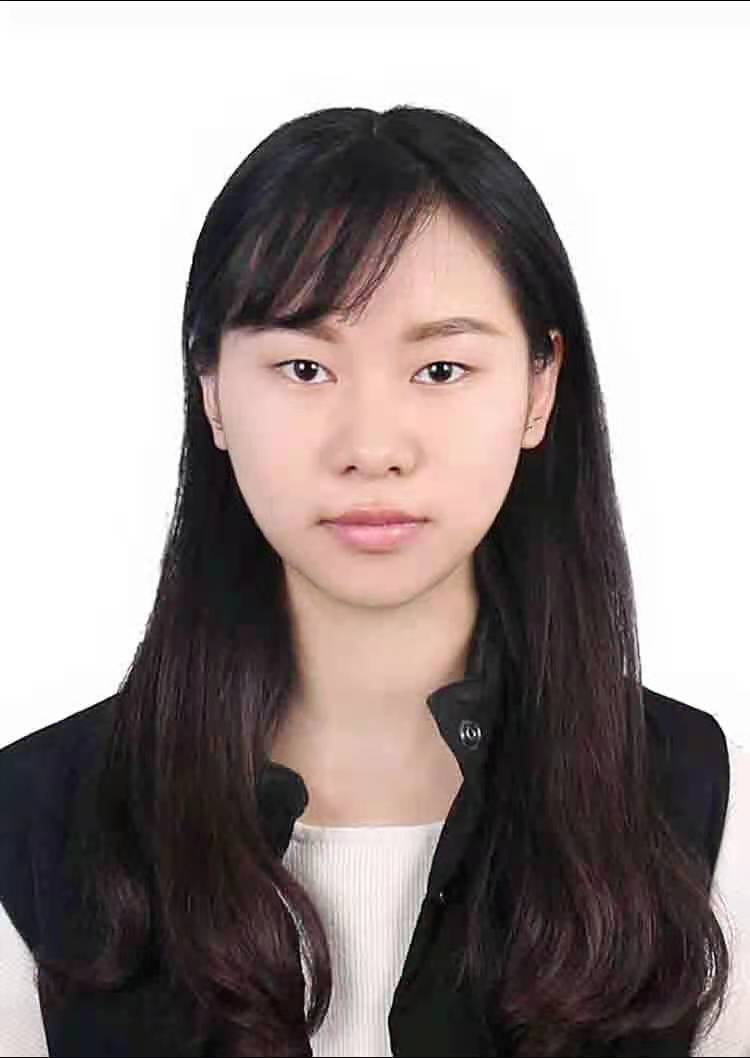}}]{Yu Huang}
received the B.E. degree from the Sichuan Normal University, Chengdu, China, in 2019. She is currently pursuing the M.S. degree in computer science in the University of Electronic Science and Technology of China (UESTC), Chengdu, China.

Her current research interests include the Internet of Things and ITS. 
\end{IEEEbiography}
\begin{IEEEbiography}[{\includegraphics[width=1in,height=1.25in,clip,keepaspectratio]{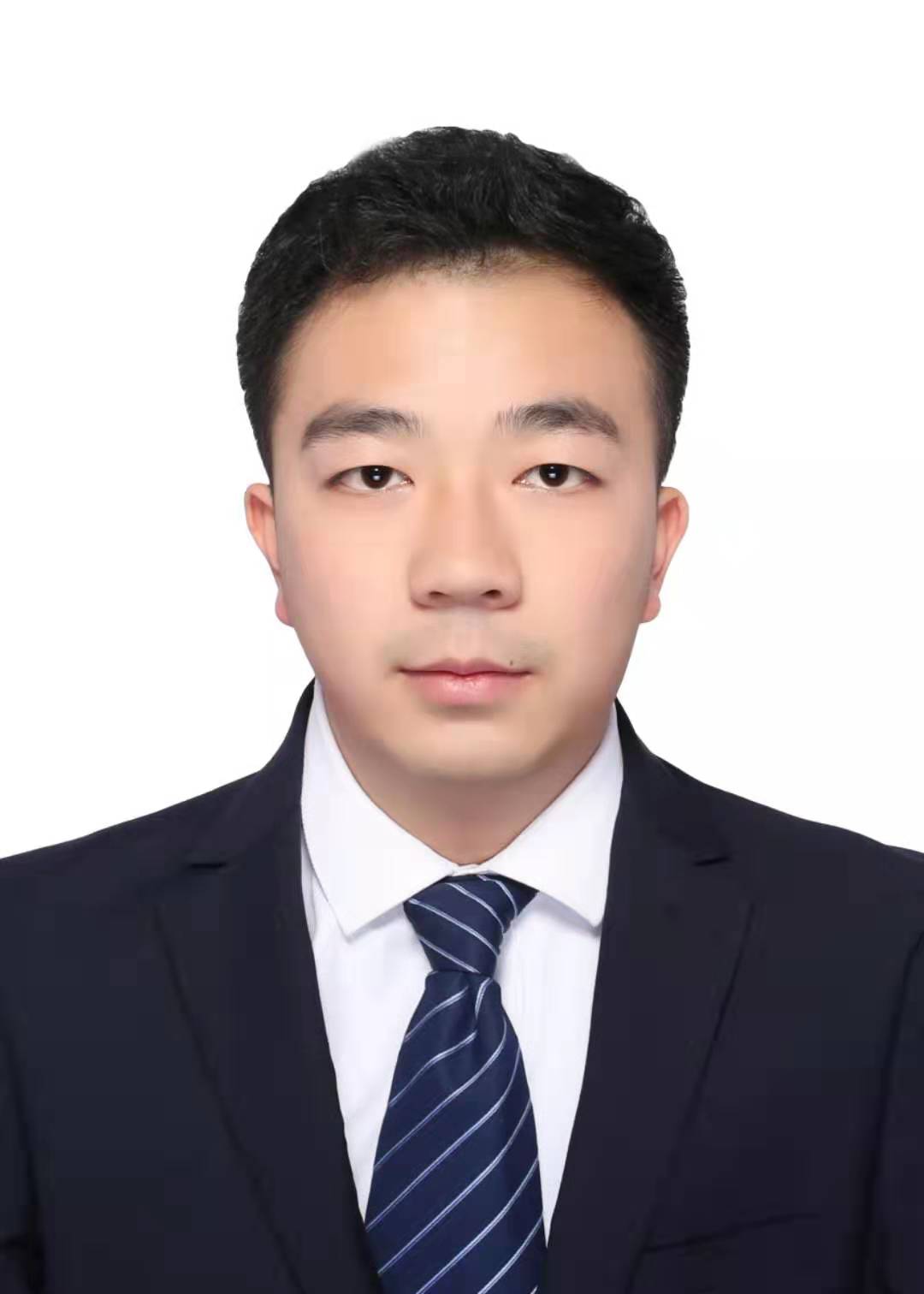}}]{Yilong Cui}
received the B.E. degree from the Lanzhou Jiaotong University, LanZhou, China, in 2017 and received the M.S. degree in computer science in the University of Electronic Science and Technology of China (UESTC), Chengdu, China in 2020.
\end{IEEEbiography}
\begin{IEEEbiography}[{\includegraphics[width=1in,height=1.25in,clip,keepaspectratio]{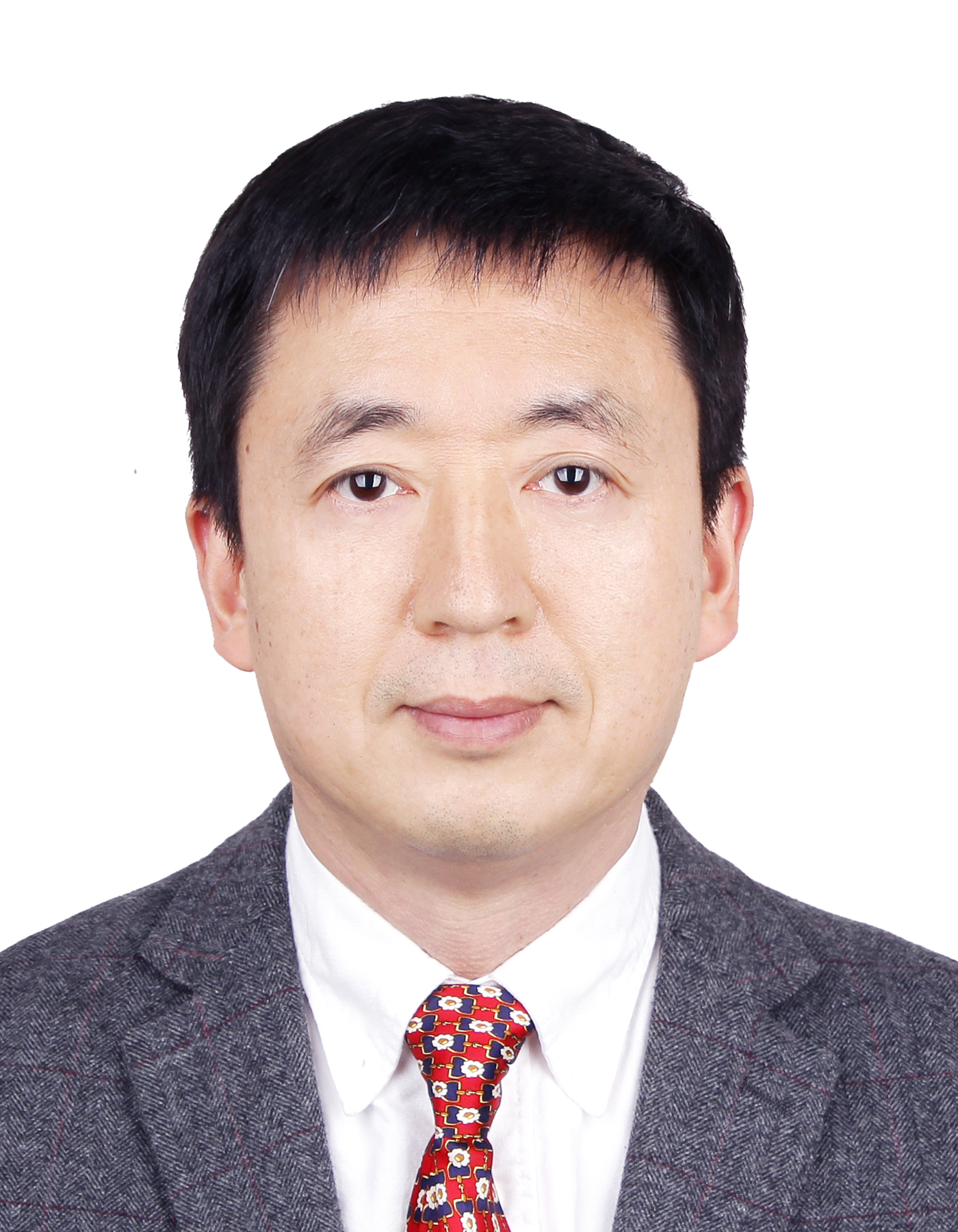}}]{Wenyong Wang}
WENYONG WANG (Member, IEEE) received the B.S. degree in computer science from Beihang University, Beijing, China, in 1988, and the M.S. and Ph.D. degrees from the University of Electronic Science and Technology (UESTC), Chengdu, China, in 1991 and 2011, respectively. He has been a Professor with the School of Computer Science and Engineering, UESTC, in 2009. He has served as the Director of the Information Center of UESTC and the Chairman of the UESTC-Dongguan Information Engineering Research Institute, from 2003 to 2009. He is currently a Visiting Professor with the Macau University of Technology. His main research interests include next-generation Internet, software-designed networks, software engineering, and artificial intelligence. He is a member of the expert board of CERNET and China Next-Generation Internet Committee and a Senior Member of the Chinese Computer Federation.
\end{IEEEbiography}
\end{document}